\documentclass[fleqn,usenatbib]{mnras}
\pdfoutput=1 %for arXiv submission
\usepackage{newtxtext,newtxmath}
\usepackage[T1]{fontenc}
\usepackage{caption}
\usepackage[font=small,labelfont=bf]{caption}
\usepackage{graphicx}	% Including figure files
\usepackage{color}
\usepackage{floatrow}
\usepackage{cancel} 
\usepackage{array}
\definecolor{purple}{rgb}{0.78,0.18,0.77}

\newcommand{\hmpc}{h^{-1}{\rm Mpc}}
\newcommand{\ihmpc}{{h} \rm{Mpc}^{-1}}

\newcommand{\tb}{T_{b}}
\newcommand{\hi}{\ion{H}{I}}
\newcommand{\tbhi}{T_{b, \ion{H}{I}}}
\newcommand{\Stb}{S \,T_{b, \ion{H}{I}}}
\newcommand{\Ntgeneral}{|S+N_{t_{\rm obs}}| \,T_{b, \ion{H}{I}}}
\newcommand{\Ntb}{|S+N_{22 \rm d}| \,T_{b, \ion{H}{I}}}

\newcommand{\FNtgeneral}{F_{k_{\rm min, \parallel}}|S+N_{t_{\rm obs}}| \, T_{b, \ion{H}{I}}}
\newcommand{\FNtb}{F_{0.3 \ihmpc}|S+N_{22\rm d}| \, T_{b, \ion{H}{I}}}
\newcommand{\FNtbA}{F^{p_2}|S+N_{22\rm d}| \, T_{b, \ion{H}{I}}}
\newcommand{\FNtbB}{F^{p_4}|S+N_{22\rm d}| \, T_{b, \ion{H}{I}}}
\newcommand{\FNdtb}{F_{0.3 \ihmpc}|S+N_{4\rm yrs}| \, T_{b, \ion{H}{I}}}
\newcommand{\FaNdtb}{F_{0.25 \ihmpc}|S+N_{4\rm yrs}| \, T_{b, \ion{H}{I}}}
\newcommand{\FbNdtb}{F_{0.2 \ihmpc}|S+N_{4\rm yrs}| \, T_{b, \ion{H}{I}}}
\newcommand{\FcNdtb}{F_{0.15 \ihmpc}|S+N_{4\rm yrs}| \, T_{b, \ion{H}{I}}}
\newcommand{\FdNdtb}{F_{0.1 \ihmpc}|S+N_{4\rm yrs}| \, T_{b, \ion{H}{I}}}
\newcommand{\FeNdtb}{F_{0.05 \ihmpc}|S+N_{4\rm yrs}| \, T_{b, \ion{H}{I}}}
\newcommand{\FNatb}{F_{0.3 \ihmpc}|S+N_{2\rm mos}| \, T_{b, \ion{H}{I}}}
\newcommand{\FNbtb}{F_{0.3 \ihmpc}|S+N_{6\rm mos}| \, T_{b, \ion{H}{I}}}
\newcommand{\FNctb}{F_{0.3 \ihmpc}|S+N_{1\rm yr}| \, T_{b, \ion{H}{I}}}

\newcommand{\CDF}{{\rm CDF}}

\newcommand{\nn}{{\rm NN}}
\newcommand{\pcfs}{{\rm PCFs}}
\newcommand{\pcf}{{\rm PCF}}
\newcommand{\random}{{R}}
\newcommand{\halo}{{H}}

\newcommand{\eq}[2]{\begin{align} \label{eq:#1} #2 \end{align}}

\title[{$\hi$-galaxy higher-order cross-correlations}]{Boosting $\hi$-Galaxy Cross-Clustering Signal through Higher-Order Cross-Correlations}

\author[Chand et al.]{
Eishica Chand $^{1}$\thanks{E-mail: {\tt eishica.chand@students.iiserpune.ac.in}},
Arka Banerjee$^{1}$\thanks{E-mail: {\tt arka@iiserpune.ac.in}},
Simon Foreman$^{2}$
and Francisco Villaescusa-Navarro$^{3,4}$\thanks{E-mail: {\tt villaescusa.francisco@gmail.com}} \\
$^{1}$Department of Physics, Indian Institute of Science Education and Research,
Homi Bhabha Road, Pashan, Pune 411008, India\\
$^{2}$Department of Physics, Arizona State University, Tempe, AZ 85287, USA\\
$^{3}$Center for Computational Astrophysics, Flatiron Institute, 162 5th Avenue, New York, NY 10010, USA\\
$^{4}$Department of Astrophysical Sciences, Princeton University, Peyton Hall, Princeton, NJ 08544, USA\\
}
\date{Accepted XXX. Received YYY; in original form ZZZ}

\pubyear{2024}

\begin{document}
\label{firstpage}
\pagerange{\pageref{firstpage}--\pageref{lastpage}}
\maketitle

\begin{abstract}
After reionization, neutral hydrogen ($\hi$) traces the large-scale structure (LSS) of the Universe, enabling $\hi$ intensity mapping (IM) to capture the LSS in 3D and constrain key cosmological parameters. We present a new framework utilizing higher-order cross-correlations to study $\hi$ clustering around galaxies, tested using real-space data from the IllustrisTNG300 simulation. This approach computes the joint distributions of $k$-nearest neighbor ($k\nn$) optical galaxies and the $\hi$ brightness temperature field smoothed at relevant scales (the $k\nn$-field framework), providing sensitivity to all higher-order cross-correlations, unlike two-point statistics. To simulate $\hi$ data from actual surveys, we add random thermal noise and apply a simple foreground cleaning model, filtering out Fourier modes of the brightness temperature field with $k_\parallel < k_{\rm min,\parallel}$. Under current levels of thermal noise and foreground cleaning, typical of a Canadian Hydrogen Intensity Mapping Experiment (CHIME)-like survey, the $\hi$-galaxy cross-correlation signal in our simulations, using the $k\nn$-field framework, is detectable at $>30\sigma$ across $r = [3,12] \, \hmpc$. In contrast, the detectability of the standard two-point correlation function ($2\pcf$) over the same scales depends strongly on the foreground filter: a sharp $k_\parallel$ filter can spuriously boost detection to $8\sigma$ due to position-space ringing, whereas a less sharp filter yields no detection.
Nonetheless, we conclude that $k\nn$-field cross-correlations are robustly detectable across a broad range of foreground filtering and thermal noise conditions, suggesting their potential for enhanced constraining power over $2\pcfs$.
\end{abstract}


%%%%%%%%%%%%%%%%%%%%%%%%%%%%%%%%%%%%%%%%%%%%%%%%%%

%%%%%%%%%%%%%%%%% BODY OF PAPER %%%%%%%%%%%%%%%%%%

%%%%%%%%%%%%%%%%%%%%%%%%%%%%%%%%%%%%%%%%%%%%%%%%%%%%%
\section{INTRODUCTION}
\label{sec:intro}
%%%%%%%%%%%%%%%%%%%%%%%%%%%%%%%%%%%%%%%%%%%%%%%%%%%%%

Neutral hydrogen ($\hi$) has been abundant in the universe since the era of baryon-photon decoupling \citep{Recombination&HI, Sunyaev2009}.
A key characteristic of $\hi$ is its hyperfine transition at a frequency of 1420 MHz, corresponding to a wavelength of 21 cm in its rest frame. This sharp line feature allows for the probing of $\hi$ across space and time. The redshifting of this line provides insights into the past conditions of $\hi$ and the processes of galaxy evolution \citep[see, e.g.,][]{HiGalaxyProbe}.
The novel $\hi$ Intensity Mapping (IM) technique  \citep{Chang_2008, Mao_2008, Wyithe2008, peterson2009, Wyithe_2009, Chang2010, Ansari2012, Battye2013, HiIM_SKA} involves mapping the total diffuse 21 cm radiation rather than resolving individual galaxies. This method captures $\hi$ emission from both bright and faint galaxies, the latter being difficult to detect in traditional optical surveys. Consequently, $\hi$ IM surveys can cover a large redshift range and a substantial cosmological volume. The resulting increase in survey volume and depth has the potential to improve constraints on various cosmological models and parameters \citep[see, e.g.,][]{Pritchard2012, HiIMStage2}. 

In the post-reionization universe ($z \lesssim 6$), $\hi$ is primarily found in the interstellar medium of galaxies \citep[see, e.g.,][]{Wolfe2005, Paco_2018}, where high density and cool temperatures enable self-shielding it from ionizing radiation. Understanding the post-reionization $\hi$ signal is crucial for cosmology because it traces galaxy centers that follow dark matter haloes, making $\hi$ itself a tracer of dark matter over-densities. The $\hi$ intensity mapping ($\hi$ IM) technique in this era allows for 3D mapping of large-scale structure (LSS) \citep[see, e.g.,][]{Bharadwaj_2001, Furlanetto2006, Chang_2008, Bharadwaj2009, Visbal_2009, Bagla_2010, Bull2015}.

Examining LSS at lower redshifts is crucial for understanding the nature of dark energy. Dark energy dominates the low-redshift energy budget of our universe and propels accelerated expansion, exerting significant influence on the growth of cosmic structures \citep[see, e.g.,][]{Riess1998, Peebles2003DE, DarkEnergyStructureGrowth}). The concordance model of cosmology \citep[see, e.g.,][]{Dodelson2003}, the $\Lambda$CDM paradigm,  assumes a cosmological constant as the component driving the universe's accelerated expansion. 
However, the recent findings from the baryonic acoustic oscillation (BAO) analysis from the dark energy spectroscopic instrument (DESI) Year One data \citep{DESI2024} are quite interesting, as they tentatively suggest a time-varying dark energy equation of state. Therefore, combining the power of $\hi$ mapping can help shed light on this question. Yet another exciting target of ongoing and upcoming LSS experiments is the search for primordial non-Gaussianity (PNG) \citep[see, e.g.,][]{Chen2010, Alvarez2014, Meerburg2019}. In this context also, the 21 cm signal holds significant potential as it uniquely traces a large part of our cosmic history and provides access to an enormous volume for probing the 3D LSS. Various studies in the literature have explored non-Gaussianity using the 21 cm signal \citep[see, e.g.,][]{Cooray2006, Pillepich2007, Crociani2009, Lidz2013, Tashiro2013, Anson2013, Li2017}.

Therefore, utilizing the full power of the $\hi$ IM technique can provide crucial inputs to answering many open questions in cosmology. However, there are challenges to fully realizing the potential of the $\hi$ IM technique. The observed $\hi$ intensity maps are typically heavily dominated by contaminants, making isolating the cosmological $\hi$ signal complex. One major obstacle in radio observations is the presence of foregrounds, which are contributed by galactic and extragalactic sources.
These foregrounds, being orders of magnitude stronger than the cosmological $\hi$ signal, completely overshadow it \citep[see, e.g.,][]{Shaver1999, Matteo2002, Oh&Mack2003, HiIMStage2}. However, these foregrounds are expected to have a different, smoother frequency dependence than the cosmological $\hi$ signal. This fact has been leveraged in various techniques proposed to clean the observed 21 cm brightness temperature field, $\tbhi$ \citep[see, e.g.,][]{FGref1, Liu:2011ih, Ansari2012, Alonso2015, Liu2020}.
Another significant source of contamination in radio observations is the thermal noise of the detector, which includes contributions from both sky and instrument temperatures. This thermal noise can limit the total signal-to-noise ratio (SNR) of the $\hi$ signal across a wide range of scales, particularly in first-generation analyses with limited volumes of data.

Given the observational challenges, detecting $\hi$ clustering through auto-correlation measurements remains difficult, although a first detection via the foreground avoidance strategy has been claimed in \citet{Paul2023}. Successful detections, however, have been achieved through cross-correlations with established LSS tracers such as galaxies \citep[see, e.g.,][]{Pen2009, Chang2010, Mausi_hiccgalaxy, Anderson_hiccgalaxy, Wolz2022, CHIME:2022kvg, Cunnington2023} and Lyman-alpha forests \citep{HI&Lymanalpha}. Cross-correlation studies benefit from mitigating survey-specific systematics, uncorrelated foregrounds, and other noise contaminants across individual surveys \citep[see, e.g.,][]{Francisco2015, Wolz2016, Pourtsidou2017}. 

After necessary instrumental calibration and foreground filtering of the observed $\tbhi$ map, several recent analyses have been primarily limited to quasi-linear to non-linear scales \citep{Paul2023,CHIME:2022kvg,HI&Lymanalpha}. This limitation is expected to improve as foreground filtering techniques become more effective, enabling the recovery of linear scales. 
Initially, fluctuations in the underlying matter density are well approximated by a Gaussian random field \citep{PlanckInflation2018}. However, non-linearities develop over time on these scales, generating non-Gaussian clustering. Traditional clustering studies rely on two-point correlation statistics, which provide a complete description of Gaussian clustering but do not capture non-Gaussian clustering information.

When studying $\hi$-galaxy cross-correlations, the $\hi$ data can be approximated as a continuous `field', while galaxy data represents a finite set of discrete `tracers'. We refer to such cross-correlations as `tracer-field' cross-correlations. The standard technique for capturing these cross-correlations has been the two-point cross-correlation functions ($2\pcfs$) or their Fourier-space counterpart, the cross-power spectrum ($P_{\mathrm{cross}}(k)$). However, as mentioned earlier, this method captures cross-clustering information only up to the second order. To extract more information from the accessible data, particularly on small scales, higher-order cross-correlations need to be exploited. Implementing such higher-order clustering techniques has the potential to significantly improve the leverage of ongoing and upcoming $\hi$ cosmological surveys.

There have been studies involving higher-order tracer-field cross-correlations between LSS tracers \citep[see, e.g.,][]{fourpointfunction, Exhigherorder1, Exhigherorder2, Doux:2016xhg, Beane:2018pmx, Chakraborty:2022aok, Farren:2023yna}, but the technique of interest here is the $k\nn$-field framework \citep{TracerField}. The core idea of this framework lies in computing the joint probabilities of finding the $k$-th nearest neighbor ($k\nn$) discrete tracer within a radial scale of interest, along with the continuous tracer field--smoothed at the same radial scale--crossing a given threshold criterion, which is a generalization of the concepts presented in \citet{kNN,kNNcross}. This joint probability is the measure of the joint clustering between the set of discrete tracers and the continuous tracer field, and dividing it by the product of marginal probabilities isolates the cross-clustering signal. 

The cross-clustering signal measured via the $k\nn$-field framework is sensitive to all higher-order cross-correlations. This framework has demonstrated high effectiveness in identifying cross-clustering patterns, even in the presence of significant noise in the continuous tracer field \citep[see, e.g.,][]{TracerField}. Another major advantage of this approach is its lower computational expense compared to other higher-order statistics. This technique has already been applied in another context, such as the spatial clustering of gravitational wave sources \citep{Kaustubh2024}.
Here, we aim to extend this framework to probe the clustering of $\hi$ around mock galaxies in simulation data.

In this paper, we demonstrate the robustness of the $k\nn$-field framework in capturing the cross-clustering between $\hi$ and mock galaxies in data from the IllustrisTNG300 simulation, particularly amidst contamination in the $\hi$ field. Section~\ref{sec:formalism} begins with a brief explanation of the theoretical formalism behind the $k\nn$-field framework, followed by an outline of the computational methods for both the $k\nn$-field framework and two-point statistics. Section~\ref{sec:simulation} describes the simulation datasets used in this study. Section~\ref{sec:hinoise} addresses observational challenges for $\hi$, including radio foregrounds and thermal noise, and details how these are incorporated into the simulated $\tbhi$ field. In Section~\ref{sec:Results}, we present the results of cross-correlations between selected haloes and the $\tbhi$ fields of interest and further explore the impact of varying contamination levels on our findings. Finally, Section~\ref{sec:summary} summarizes the main conclusions and offers a brief outlook on future directions.

%%%%%%%%%%%%%%%%%%%%%%%%%%%%%%%%%%%%%%%%%%%%%%%%%%
\section{FORMALISM AND COMPUTATION}
\label{sec:formalism}
%%%%%%%%%%%%%%%%%%%%%%%%%%%%%%%%%%%%%%%%%%%%%%%%%%

In this section, we briefly explain the theoretical formalism of the $k\nn$-field framework, as originally outlined in \citet{TracerField}. Subsequently, we elaborate on the methodology employed for computing tracer-field cross-correlations using both the $k\nn$-field framework and the $2 \pcfs$. 

%%%%%%%%%%%%%%%%%%%%%%%%%%%%%%%%%%%%%%%%%%%%%%%%%%
\subsection{$k\nn$-Field framework}
\label{sec:tracerfield_framework}
%%%%%%%%%%%%%%%%%%%%%%%%%%%%%%%%%%%%%%%%%%%%%%%%%%
\begin{figure*}        
\includegraphics[width=0.9\textwidth, trim = 25 0 15 0]{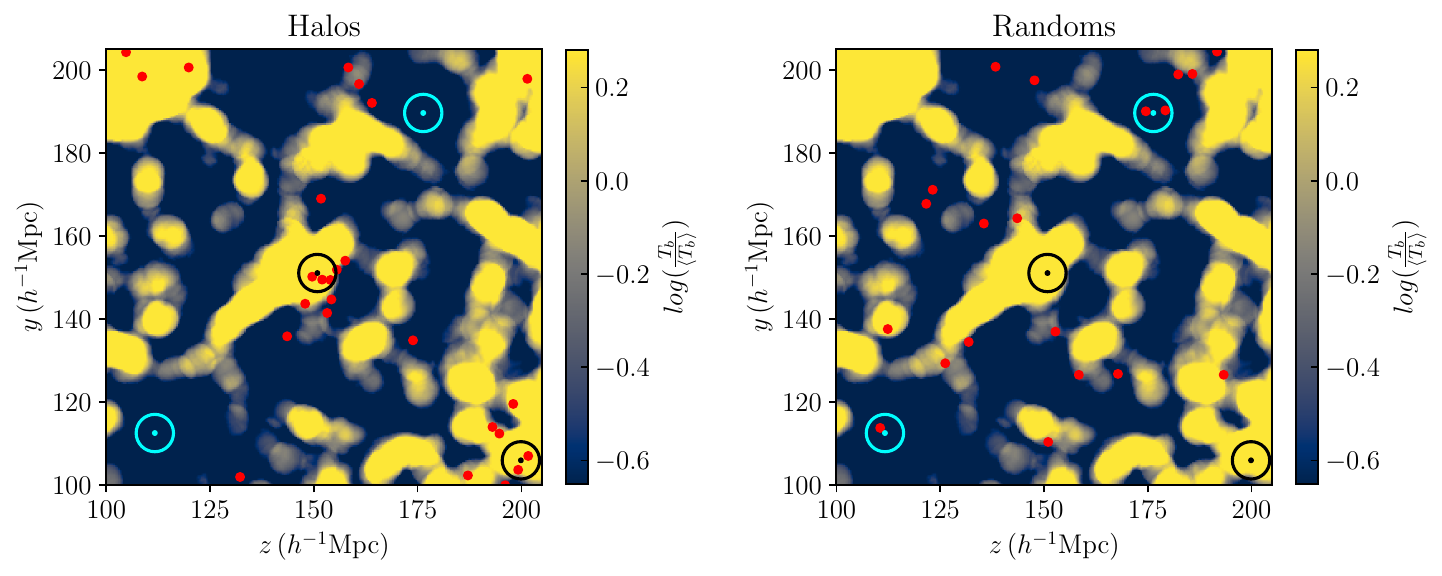}
   \caption{Comparison of halos vs. random points clustering around the $\hi$ field:
    A zoomed-in view of the IllustrisTNG300 simulation box ($L = 205 \, \hmpc$), projected over $5 \, \hmpc$ along the $x$-axis, with the $y$ and $z$ axes spanning from $100 \, \hmpc$ to $205 \, \hmpc$. The underlying field shows $\log(T_b / \langle T_b \rangle)$, smoothed over $4.5 \, \hmpc$. Two black points mark query points in high-$\tbhi$ regions (yellow), each surrounded by a $4.5 \, \hmpc$ radius circle. Similarly, two cyan points indicate query points in low-$\tbhi$ regions (navy), with corresponding circles. In the left panel, red circles represent halos in the projected region among the 3500 most massive halos. In contrast, in the right panel, red circles denote points from a set of 3500 randomly selected points within the same region. As expected, query points in high-$\tbhi$ regions generally have a higher probability of finding halos within the selected radius, while those in low-$\tbhi$ regions have a lower probability. In contrast, randoms may, by chance, fall within the radius of low-$\tbhi$ query points, and high-$\tbhi$ query points may not always find halos within the radius.}
    \label{fig:ToyExTb}
\end{figure*}

The $k\nn$-field framework unifies the discrete and continuous versions of nearest-neighbor cumulative distribution function (CDF) measurements, applied to discrete and continuous tracers, respectively, to quantify the cross-clustering between these two sets of tracers.
\begin{itemize}
    \item The $k\nn$ statistics, originally developed by \citet{kNN} to characterize the clustering of discrete tracers of the matter field at all orders, is based on the probability of finding $\geq k$ discrete tracers within a radius $r$, denoted by $\mathcal{P}_{\geq k}(r)$. In a volume, the CDF of the $k^{\rm th}$ nearest-neighbor distance from points densely sampling the full volume$-$namely, query points$-$follows:
    \eq{pdfcdf}{\mathcal{P}_{\geq k}(r) = \text{CDF}_{k\nn}(r).}
    This probability, $\mathcal{P}_{\geq k}$, depends on all connected $N$-point correlation functions, $\xi^{(N)}$, of the underlying continuous field \citep[see, e.g.,][]{White1979, Szaupdi1993, kNN}, rather than being restricted to second-order correlations (power spectrum) or third-order correlations (bispectrum), and is rather fast to compute. The integral expression for $\mathcal{P}_{\geq k}$ in terms of $\xi^{(N)}$ is detailed in \citet{kNN}. \citet{Wang2022} conducted the first study applying the $k\nn$ statistics to observational data, demonstrating its robustness over $2\pcfs$ in detecting clustering in the 1000 richest redMaPPer clusters (SDSS DR8 catalog). In a separate study, \citet{YuanTom2023} showed that 2D $k\nn$ statistics, in addition to being computationally efficient, also provide significantly tighter constraints on the galaxy-halo model than redshift-space 2$\pcfs$.

    \item The extension of the discrete $k\nn$ framework to a continuous field (e.g., the density field) was formalized in \citet{TracerField}. In the limit where a continuous field is traced by an infinitely large number of discrete tracers, the quantity mapped to a specific $k$ value for discrete tracers (as discussed previously) at a given radius $r$ corresponds to a threshold value of the field smoothed on the scale $r$: $k = \bar{n} \left(\frac{4}{3} \pi r^3\right) \rho^*_r$, where $\bar{n}$ is the number density of tracers and $\rho^*_r$ represents the density threshold of the field. As shown in \citet{TracerField}, $\mathcal{P}_{k}(r) \propto \mathrm{PDF}(\rho^*_r)$, where PDF refers to the probability density function, and $\mathcal{P}_{\geq k}(r) \propto 1 - \mathrm{CDF}(\rho^*_r)$. Like the discrete $k\nn$ CDFs, which encapsulate sensitivity to all $N$-point correlations, the PDF and CDF of the continuous field are similarly influenced by all higher-order correlations \citep[see, e.g.,][]{Bernardeau2014, Friedrich2018, Uhlemann2020}.
    
\end{itemize}

In practice, the $k\nn$-field framework relies on the simultaneous application of the discrete and continuous nearest-neighbor CDF measurements described above. Its core concept can be qualitatively understood from Fig.~\ref{fig:ToyExTb}, constructed using IllustrisTNG300 simulation datasets, the $\tbhi$ field, and dark matter halo catalogue. The IllustrisTNG project \citep{Annalisa2018, Dylan2018, Volker2018, Marinacci2018, Naiman2018, IllustrisTNG} is a state-of-the-art cosmological magneto-hydrodynamic simulation of galaxy formation with TNG300 component having a side length of $300$ Mpc ($205 \,\hmpc$).

The figure presents a zoomed-in view of a subregion within the IllustrisTNG300 simulation box. This subregion is projected along the $x$-axis with a thickness of $5 \, \hmpc$, while the $y$ and $z$ coordinates span the range $100 \, \hmpc$ to $205 \, \hmpc$. The background field in both panels represents $\log(T_{b}/\langle T_{b} \rangle)$, where $T_{b}$ denotes the $\hi$ brightness temperature, smoothed using a real-space top-hat kernel with a radius of $4.5 \, \hmpc$.

In the $k\nn$ framework, the cosmological box is densely sampled by selecting a large number of query points$-$either placed randomly or arranged on a grid, with random placement used in this case, such that their number exceeds that of the data points. The figure highlights four such query points: two points chosen in high-$\tbhi$ regions (marked in black) and two in low-$\tbhi$ regions (marked in cyan), each enclosed by a circle of radius $4.5 \, \hmpc$. The left panel represents the Halos case, where red circles indicate the locations of halos among the 3500 most massive halos in the projected region. The right panel corresponds to the Randoms case, where red circles denote points randomly selected from a set of 3500 points within the same region.

From the figure, one can see that the probability of finding, say, at least one halo ($k=1$) within a radius of $4.5 \, \hmpc$ is high for a high-$\tbhi$ query point and low (or zero, in this case) for a low-$\tbhi$ query point. This occurs because halos cluster with the $\hi$ field, as both trace the same underlying density field. In contrast, the right panel illustrates that when there is no intrinsic correlation between the underlying field and discrete tracers, a low-$\tbhi$ query point may still have a high probability, and a high-$\tbhi$ query point may have a low probability of finding at least one random point within the selected radius$-$purely by chance. Averaging over many such realizations of random positions would yield a zero cross-clustering signal.

To further illustrate the framework quantitatively, consider two datasets within a cosmological volume. The first dataset consists of discrete tracers, represented as a collection of points whose locations follow a local Poisson process governed by a continuous field with density $\tilde{\rho}(x)$. The number density of these points is denoted by $\bar{n}$. The second dataset represents a continuous tracer field with density $\rho(x)$. We omit the explicit spatial dependence ($x$) for brevity in our notation for densities. For visualization, Fig.~\ref{fig:ToyExTb} can be referred to: the red circles denote the discrete tracers, while the underlying colormap represents the continuous field, both coexisting within the same volume.

To assess the cross-clustering between these two tracers using the $k\nn$-field framework, we compute the joint probability of finding at least $k$ discrete data points within a sphere of radius $r$ and the density of the continuous tracer field, smoothed over the same radius, exceeding a threshold $\rho_r^*$. This joint probability is denoted as $\mathcal{P}_{\geq k, >\rho_r^*}$. For a given $k$ and chosen $\rho_r^*$, we express this as:
\eq{joint_CDF_def}{\mathcal{P}_{\geq k, > \rho_r^*} = \mathcal{P}_{>\rho_r^*} - \sum_{k^\prime<k}\mathcal{P}_{k^\prime, >\rho_r^*},}
where $\mathcal{P}_{>\rho_r^*}$ represents the marginal probability of the continuous field smoothed over the sphere of radius $r$ crossing above the density threshold $\rho_r^*$, and, 
\eq{joint_prob_def}{\mathcal{P}_{k, >\rho_r^*} =  \int_{\rho_r^*}^{\infty} \int_{0}^{\infty} \frac{\left(\lambda(\tilde \rho_r)\right)^k}{k!} e^{-\lambda(\tilde \rho_r)} \phi(\tilde \rho_r, \rho_r)\, \mathrm{d}\tilde \rho_r \, \mathrm{d}\rho_r,}
where,
\eq{lambda}{\lambda(\tilde \rho_r) = \bar n \, \Bigg (\frac{4 \pi r^3}{3}\Bigg) \Bigg(\frac{\tilde \rho_r}{\langle \tilde \rho_r\rangle}\Bigg),}
and $\phi(\tilde \rho_r,  \rho_r)$ represents the joint probability distribution of the two fields $\tilde{\rho}$ and $\rho$ when smoothed on scale $r$. It captures the complete information about the correlations between the fluctuations of the two fields. Thus, the joint probability measure defined by Eq.~\ref{eq:joint_CDF_def} is sensitive to correlations present in the fluctuations between the two kinds of tracers at all orders.

If the two fields (densities $\tilde \rho$ and $\rho$) under consideration are statistically independent, their joint probability distribution simplifies to the product of individual probability distributions, i.e., $\phi(\tilde \rho_r, \rho_r) \propto \phi_1(\tilde \rho_r) \phi_2(\rho_r)$. This implies for the joint probability: 
\eq{joint_CDF_product}{\mathcal P_{\geq k, >\rho_r^*} = \mathcal P_{\geq k} \times \mathcal P_{>\rho_r^*} \, .} 
We use the ratio $\mathcal P_{\geq k, >\rho_r^*} / (\mathcal P_{\geq k} \times \mathcal P_{>\rho_r^*})$ as a measure of cross-clustering between the two fields, beyond their auto-clustering, capturing all higher-order cross-correlations between the continuous tracer field and the set of discrete tracers.

Previous studies have demonstrated the relative performance of $k\nn$ CDFs over the combination of second and third-order statistics in other contexts. For instance, \citet{DES2023} showed that for the galaxy lensing convergence field, CDF measurements provide better cosmological constraints compared to the combination of second and third-order moments. Similarly, \citet{Coulton2024} demonstrated that $k\nn$ CDFs outperform the combined power spectrum and bispectrum in identifying small-scale signatures of primordial non-Gaussianity (PNG), especially when marginalizing over uncertainties in the galaxy-halo connection.

%%%%%%%%%%%%%%%%%%%%%%%%%%%%%%%%%%%%%%%%%%%%%%%%%%
\subsection{Computation using $k\nn$-Field Framework}
\label{sec:$k$NNfcompute}
%%%%%%%%%%%%%%%%%%%%%%%%%%%%%%%%%%%%%%%%%%%%%%%%%%

To compute the cross-correlation signal defined via the $k\nn$-field framework between two given datasets, a set of discrete data points and a continuous field, the following procedure is employed (we refer the reader to \citet{TracerField} for detailed information):
\begin{enumerate}
    \item Define $N_{q}^3$ query points distributed randomly in the cosmological volume of box-size $L$, ensuring that $N_{q}^3$ is much larger than the number $N$ of discrete data points.
    \item Construct a $k$-d tree from the set of data points. Use this tree to measure the distance to each query point's $k$-nearest neighbor data points. For each value of $k$, sort these distances to obtain the empirical cumulative distribution function, $\CDF_{k\nn}(r)$, which approaches $\mathcal P_{\geq k}(r)$ in the limit of large $N_q^3$.
    \item The continuous tracer field is defined on a grid with $N_{g}^3$ grid points. First, we smooth this field using a spherical top-hat function with a radius $r$, and then we interpolate it to the positions of $N_{q}^3$ query points defined in the initial step. We denote the density of this smoothed field as $\rho_r$.
    \item Choose the density threshold value, $\rho_r^*$, as a fixed percentile of the density values $\rho_r$ across all values of $r$. \footnote{We will present results using the $75$th percentile of smoothed densities as the threshold in this study.}
    \item Calculate the fraction of $N_q^3$ query points for which the interpolated smoothed field exceeds $\rho_r^*$. This yields an empirical measure of $\mathcal{P}_{>\rho_r^*}$.
    \item Compute the fraction of $N_q^3$ query points that locate their $k$-th nearest neighbor data point within a distance of $\leq r$, while simultaneously having the interpolated smoothed density field at these query positions surpass the density threshold $\rho_r^*$. In the limit of large $N_q^3$, this yields $\mathcal{P}_{\geq k, > \rho_r^*}$.

\end{enumerate}
The quantity $\mathcal{P}_{\geq k, >\rho_r^*}/(\mathcal{P}_{\geq k} \times \mathcal{P}_{>\rho_r^*})$, which we refer to as the Excess-CDF, measures the cross-correlation signal between the discrete tracers and the continuous tracer field, independent of their auto-clustering contributions.

\begin{table*}
\captionsetup{width=0.85\textwidth} % Centering the caption and setting its width

\renewcommand{\arraystretch}{1.4} % Adjust this value for more or less space
\begin{tabular}{|p{3cm}|p{9cm}|}
    \hline
    \textbf{Notation} & \textbf{Description} \\ 
    \hline
    $\Stb$ & The original real-space 21 cm $\tb$ field obtained from the IllustrisTNG $300$ simulation at a redshift $z = 1$. The underlying field depicted in Fig.~\ref{fig:STb} illustrates $\Stb$. \\ 
    \hline
    $\Ntb$ & The 21 cm $\tb$ field generated after adding thermal noise, corresponding to $22$ days of CHIME's observation, to the original field, $\Stb$. The underlying field in the first panel of Fig.~\ref{fig:Tb_FNcombine} illustrates $\Ntb$. \\ 
    \hline
    $\FNtbA$ & The 21 cm $\tb$ field obtained after performing foreground filtering with filter $F^{p_2}: 1 - \exp{\left(-k_{\parallel}^2 / {2k_{\rm min, \parallel}^2}\right)}$, where $k_{\rm min, \parallel} = 0.3\, \ihmpc$, on the noise-contaminated field, $\Ntb$. The underlying field in the second panel of Fig.~\ref{fig:Tb_FNcombine} illustrates $\FNtbA$. \\ 
    \hline
    $\FNtbB$ & The 21 cm $\tb$ field obtained after performing foreground filtering with filter $F^{p_4}: 1 - \exp{\left(-k_{\parallel}^4 / {2k_{\rm min, \parallel}^4}\right)}$, where $k_{\rm min, \parallel} = 0.3 \,\ihmpc$, on the noise-contaminated field, $\Ntb$. The underlying field in the third panel of Fig.~\ref{fig:Tb_FNcombine} illustrates $\FNtbB$. \\ 
    \hline 
\end{tabular}

\caption{Notation conventions for the simulated $\tb$ fields. $\Stb$ represents the original simulated 21 cm field. Adding thermal noise to $\Stb$, corresponding to 22 days of CHIME observation, results in $\Ntb$. Applying a Gaussian-form foreground filter to $\Ntb$ yields $\FNtbA$, while a sharper filter (with a power of 4 instead of 2 in the exponent) produces $\FNtbB$. Both filters use a minimum wavenumber of $k_{\rm min, \parallel} = 0.3 \, \ihmpc$, except in Sec.~\ref{sec:varyf&n}, where this cutoff is varied.}
\label{tab:tbfields}
\end{table*}

%%%%%%%%%%%%%%%%%%%%%%%%%%%%%%%%%%%%%%%%%%%%%%%%%%
\subsection{Computation using Two-Point Statistics}
\label{sec:twoptcompute}
%%%%%%%%%%%%%%%%%%%%%%%%%%%%%%%%%%%%%%%%%%%%%%%%%%

To compute the two-point cross-correlation between a set of discrete data points and a continuous field, we stack the field into spherical shells of radius $r$ and thickness $dr$ around the positions of the discrete tracers.\footnote{The chosen thickness, $dr$, is $1 \,\hmpc$ ($\pm\, 0.5 \, \hmpc$).} The enclosed density is then averaged over all data points, yielding the two-point cross-correlation measure at scale $r$ between the data points and the field, following a similar procedure to \citet{TracerField}.

It's crucial to emphasize that our analysis primarily focuses on $\tbhi$, which directly correlates with the $\hi$ density field \citep{CHIME:2022kvg}. As a result, the density variable $\rho$ mentioned previously and in Sec.~\ref{sec:$k$NNfcompute} will be replaced with $\tbhi$, representing the 21 cm brightness temperature field. Additionally, for our analysis, the mentioned discrete tracer will correspond to halo positions within the simulation box.

%%%%%%%%%%%%%%%%%%%%%%%%%%%%%%%%%%%%%%%%%%%%%%%%%%
\section{SIMULATION DATA}
\label{sec:simulation}
%%%%%%%%%%%%%%%%%%%%%%%%%%%%%%%%%%%%%%%%%%%%%%%%%%

\begin{figure}
    \includegraphics[width=0.9\textwidth, trim=0 30 0 15]{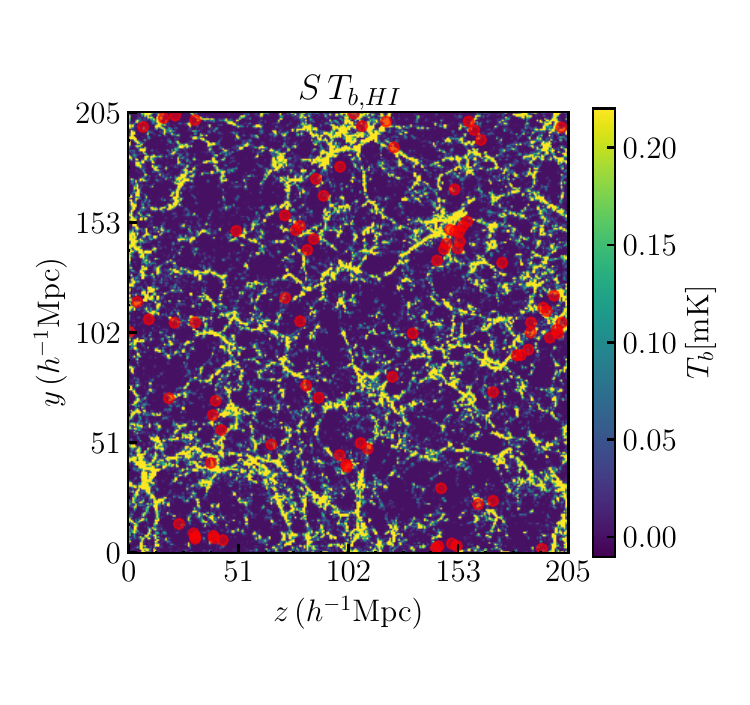}
    \caption{Visualization of a $5 \, \hmpc$ projected region along the $x$-axis of the IllustrisTNG300 simulation box, which has a full size of $L = 205 \, \hmpc$. The red circles indicate the positions of selected haloes that lie within the projected region (see Sec.~\ref{sec:simulation}). The underlying field represents the original simulation brightness temperature field, denoted by $\Stb$. Yellow regions indicate areas of high $\hi$ densities, around which most haloes are concentrated, while purple regions represent areas with sparse $\hi$, containing very few haloes. The results of cross-correlating $\Stb$ with the $4000$ most massive haloes (without foreground or thermal noise) are presented in Appendix \ref{sec:a1}.}
    \label{fig:STb}
\end{figure}

We utilized real-space IllustrisTNG300 simulation data for our analysis. The datasets of interest from the simulation are the dark matter host halo catalogs associated with galaxies and the 21 cm brightness temperature field denoted by $\Stb$, at a redshift of $z = 1$, computed following the prescription from \citet{Paco_2018}.
This redshift choice aligns with the operational range of the Canadian Hydrogen Intensity Mapping Experiment (CHIME) and falls within the observed redshift range of advanced optical surveys such as eBOSS \citep{Bolton2012, Dawson2013} and DESI \citep{DESI2016a}. These surveys target various LSS tracers, each with different number densities across redshifts. 

At a redshift of $z = 1$, Emission Line Galaxies (ELGs), characterized by their strong emission lines in the optical spectrum \citep[see, e.g.,][]{Gonzalez2020, Yuan2023}, exhibit the highest expected number density. Typically, for ELGs, this value is $n \approx 4.6 \times 10^{-4} \, [\hmpc]^{-3}$ \citep{Alam2021}. We calculated the corresponding total number of discrete tracers to match this estimated ELG number density within our simulation volume. Given the TNG300 simulation box volume, $V = 205^3 \, [\hmpc]^3$, applying the relation $N = nV$ yields $N \approx 4000$. We chose to select the 4000 most massive haloes from the complete simulated halo catalogs for our study. While a more careful analysis could have been conducted by specifically selecting haloes that are potential hosts for ELGs, we opted for this crude selection due to the exploratory nature of our study, aimed at testing the robustness of the $k\nn$-field framework.

In Fig.~\ref{fig:STb}, we present a visualization of the original simulation datasets. The background field shows a $5 \, \hmpc$ projection along the $x$-axis of $\Stb$. Red circles indicate the halo positions, that lie within this projected region. Yellow regions correspond to higher values of $\Stb$, representing higher underlying $\hi$ densities, while purple regions indicate lower $\hi$ density. The clustering of red circles near high $\hi$ density regions and their sparsity in low $\hi$ density regions aligns with the expectation that both datasets trace fluctuations in the same underlying dark matter field. In observational data, however, the presence of contaminants is expected to result in a noisy $\tbhi$ field. For validation, we also examine the cross-correlation of fluctuations in $\Stb$ with the selected haloes, with results provided in Appendix \ref{sec:a1}.

\begin{figure*}        
\includegraphics[width=1\textwidth, trim = 0 30 0 15]{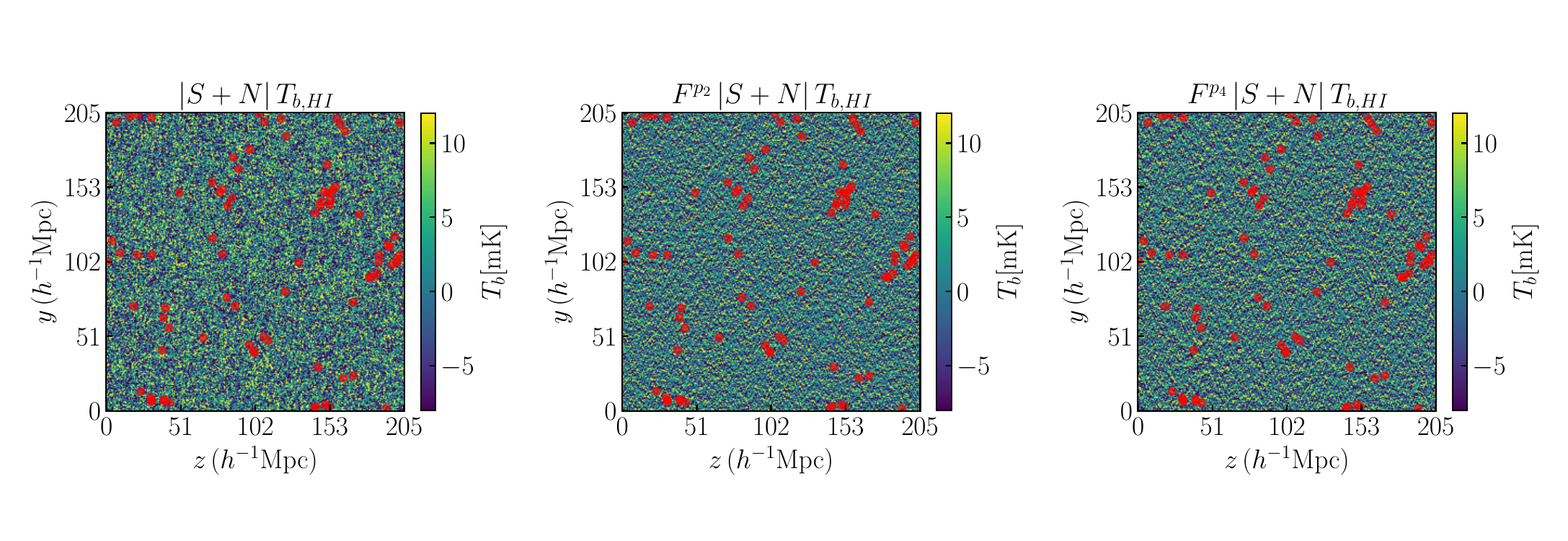}
   \caption{Visualization of a $5 \, \hmpc$ projected region along the $x$-axis of the IllustrisTNG300 simulation box. The underlying field in the \textit{first panel} shows $\Ntb$, the 21 cm brightness temperature field generated by adding thermal noise contamination to the original field, $\Stb$, based on CHIME's thermal noise model for a $22$-day observation period (see Sec.~\ref{sec:ThNoise}). The thermal noise dominates, obscuring $\hi$ fluctuations and their clustering with haloes (red circles). The underlying field in the \textit{second panel} depicts $\FNtbA$, the brightness temperature field after applying the $k$-space Gaussian filter $F^{p_2}$ to $\Ntb$, while the \textit{third panel} shows $\FNtbB$, obtained by applying the sharper $F^{p_4}$ filter on $\Ntb$ (see text for details). These filters simulate the effects of foreground filtering (see Sec.~\ref{sec:FGs}); however, the filtering effects are not prominent due to the dominance of thermal noise. Cross-correlation measures between haloes and the filtered fields are presented in Sec.~\ref{sec:Results}.}
    \label{fig:Tb_FNcombine}
\end{figure*}

%%%%%%%%%%%%%%%%%%%%%%%%%%%%%%%%%%%%%%%%%%%%%%%%%%
\section{$\hi$ Survey Contaminants}
\label{sec:hinoise}
%%%%%%%%%%%%%%%%%%%%%%%%%%%%%%%%%%%%%%%%%%%%%%%%%%

%%%%%%%%%%%%%%%%%%%%%%%%%%%%%%%%%%%%%%%%%%%%%%%%%%
\subsection{Thermal Noise}
\label{sec:ThNoise}
%%%%%%%%%%%%%%%%%%%%%%%%%%%%%%%%%%%%%%%%%%%%%%%%%%

A major source of contamination in radio observations of the cosmological $\hi$ signal is the thermal noise from detectors. The root-mean-square (RMS) of this noise depends on the system temperature, which is the sum of the sky temperature, $T_{\rm sky}$, and the instrument temperature, $T_{\rm inst}$, such that $T_{\rm sys} = T_{\rm sky} + T_{\rm inst}$. For radio telescopes, this thermal noise component is typically modeled as uncorrelated Gaussian noise \citep[see, e.g.,][]{Wilson2009,Bull2015}.

We simulate the impact of thermal noise contamination on the original simulated $\hi$ brightness temperature field, to investigate its effect on the measurement of the cross-correlation signal between $\hi$ and galaxies. To achieve this, we consider the noise covariance model of the CHIME \citep[see Appendix A.3 of][]{CHIMEoverview}. Using the specifications of CHIME and $22$ days of clean observation time, we determine the noise power spectrum corresponding to the $\hi$ density as $P_N \approx 112603 \, (\hmpc)^3$. We select $22$ days because this represents the total integration time for CHIME's first cross-correlation measurements with the eBOSS survey \citep{CHIME:2022kvg}; later, in Sec.~\ref{sec:varyf&n}, we also analyze fields with lower noise levels achievable after longer integration times. Using this model, we generated the thermal noise field on a grid matching the size of $\Stb$.
Subsequently, we added the generated thermal noise field to $\Stb$, resulting in a field referred to as $\Ntb$. It is important to note that $\Ntb$ no longer represents a physical field bounded below by 0; instead, it can take negative values. 
The first panel in Fig.~\ref{fig:Tb_FNcombine} provides a visualization of $\Ntb$, including details of the projected region and red circles, similar to those in Fig.~\ref{fig:STb}. From the figure, it is evident that the thermal noise dominates over the cosmological $\hi$ signal, with no discernible $\hi$ fluctuations. We further address the effects of foreground filtering on $\Ntb$ in Sec.~\ref{sec:FGs}.

%%%%%%%%%%%%%%%%%%%%%%%%%%%%%%%%%%%%%%%%%%%%%%%%%%%%
\subsection{Foregrounds}
\label{sec:FGs}
%%%%%%%%%%%%%%%%%%%%%%%%%%%%%%%%%%%%%%%%%%%%%%%%%%%%

In radio observations, foregrounds pose a major challenge in extracting the true cosmological 21 cm signal from the observed $\tbhi$ field. Many foreground components in the sky exhibit smooth spectral variations \citep{FGref1, FGref3, FGref2}. As a result, when analyzing the field in Fourier space of the cosmological cube, foregrounds predominantly contaminate the low $k_{\parallel}$ modes. As a rough approximation to the effects of foreground contamination on real data, one can define a minimum $k_{\parallel}$ value, $k_{\rm min, \parallel}$, below which foregrounds dominate and above which the observations are assumed to be foreground free. This minimum $k_{\parallel}$ mode determines a corresponding minimum value of $k$ through the relation $k^2 = k_{\perp}^2 + k_{\parallel}^2$. The precise value of $k_{\rm min, \parallel}$ depends on the level of foreground cleaning achievable in a given survey.

To account for the impact of foregrounds on our analysis of $\hi$ clustering around galaxies, we apply a foreground filtering procedure to the $\Ntb$ field. The observed $\tb$ field is a combination of the cosmological $\hi$ signal, contaminated by foregrounds and thermal noise from the instrument. Our filtering assumes that foregrounds completely obscure modes with $k_{\parallel} < k_{\rm min, \parallel}$ in the simulation field $\Stb$. We employ two filters: a smooth Gaussian filter, $F^{p_2} = 1 - \exp{\left(-k_{\parallel}^2 / {2k_{\rm min, \parallel}^2}\right)}$, and a comparatively sharper filter, $F^{p_4} = 1 - \exp{\left(-k_{\parallel}^4 / {2k_{\rm min, \parallel}^4}\right)}$. These filters effectively reduce the contribution of obscured modes to near-zero in $\Ntb$. We choose the value of $k_{\rm min, \parallel} = 0.3\, \ihmpc$, roughly matching the cutoff used in \citet{CHIME:2022kvg}.
We denote the filtered fields obtained from this procedure by $\FNtbA$ and $\FNtbB$. Later, in Sec.~\ref{sec:varyf&n} we vary our filter choices by using lower $k_{\rm min, \parallel}$ values to consider the improved level of foreground cleaning achievable in future surveys.

The underlying fields in the second and third panels of Fig.~\ref{fig:Tb_FNcombine} illustrate the filtered fields. The details of the projected region and red circles are consistent with those in the first panel and Fig.~\ref{fig:STb}. However, the effects of filtering are not discernible due to the dominance of thermal noise, which exhibits a white spectrum. As a result, removing a few low-$k$ modes from the $\Ntb$ field only marginally reduces the noise power.
The contaminated fields thus obtained mimic the typical 21 cm $\tb$ field accessible (after filtering) via recent $\hi$ observations. We analyze the cross-correlations of these fields with the 4000 most massive haloes from simulations and present the results in Sec.~\ref{sec:Results}.

Table~\ref{tab:tbfields} summarizes the various $\tbhi$ fields utilized in this study. The first column presents the symbols representing the $\tbhi$ fields, while the second column describes the corresponding fields.

%%%%%%%%%%%%%%%%%%%%%%%%%%%%%%%%%%%%%%%%%%%%%%%%%%
\section{ANALYSIS AND RESULTS}
\label{sec:Results}
%%%%%%%%%%%%%%%%%%%%%%%%%%%%%%%%%%%%%%%%%%%%%%%%%%

In this section, we present the main results from our analysis of the full 3D cross-correlation between the 4000 most massive haloes and the 21 cm brightness temperature field, with both datasets derived from the IllustrisTNG300 simulation at redshift $z = 1$ in real space. The simulated 21 cm brightness temperature field, $\Stb$, has been processed to account for thermal noise contamination and foreground filtering, resulting in the fields $\FNtbA$ and $\FNtbB$ (described in Sec.~\ref{sec:hinoise}), which are then analyzed for cross-correlation with haloes.

%%%%%%%%%%%%%%%%%%%%%%%%%%%%%%%%%%%%%%%%%%%%%%%%%%
\subsection{Cross-Correlation Measurements}
\label{sec:crosscorreln}
%%%%%%%%%%%%%%%%%%%%%%%%%%%%%%%%%%%%%%%%%%%%%%%%%%

The cross-correlation between selected haloes and the $\tbhi$ field of interest is analyzed using the respective cross-correlation analysis pipelines for the $k\nn$-field framework and the two-point cross-correlations. Specifically, we utilize the $1\nn$ Excess-CDF measurements computed via the $k\nn$-field framework (explained in Sec.~\ref{sec:$k$NNfcompute}) for this study. For computing the $2\pcfs$, we employ the procedure of stacking the field over halo positions (as detailed in Sec.~\ref{sec:twoptcompute}).

Given that the number of haloes in a locally observed region follows a Poisson distribution, its variance introduces a shot noise term into correlation measurements. To quantify the effects of this noise arising from the finite number of haloes (4000 in our case) on the measured cross-correlation, we perform the cross-correlation measurement of the $\tbhi$ field of interest with 2000 different realizations of 4000 randomly sampled (uncorrelated) points. Ideally, each random realization of points should yield zero cross-correlation with the $\hi$ fluctuations, indicating no correlation. However, deviations from zero in the actual computations provide a measure of the noise term associated with the finite set of haloes.

\begin{figure}
    \includegraphics[width= 1\textwidth, trim = 25 0 15 0]{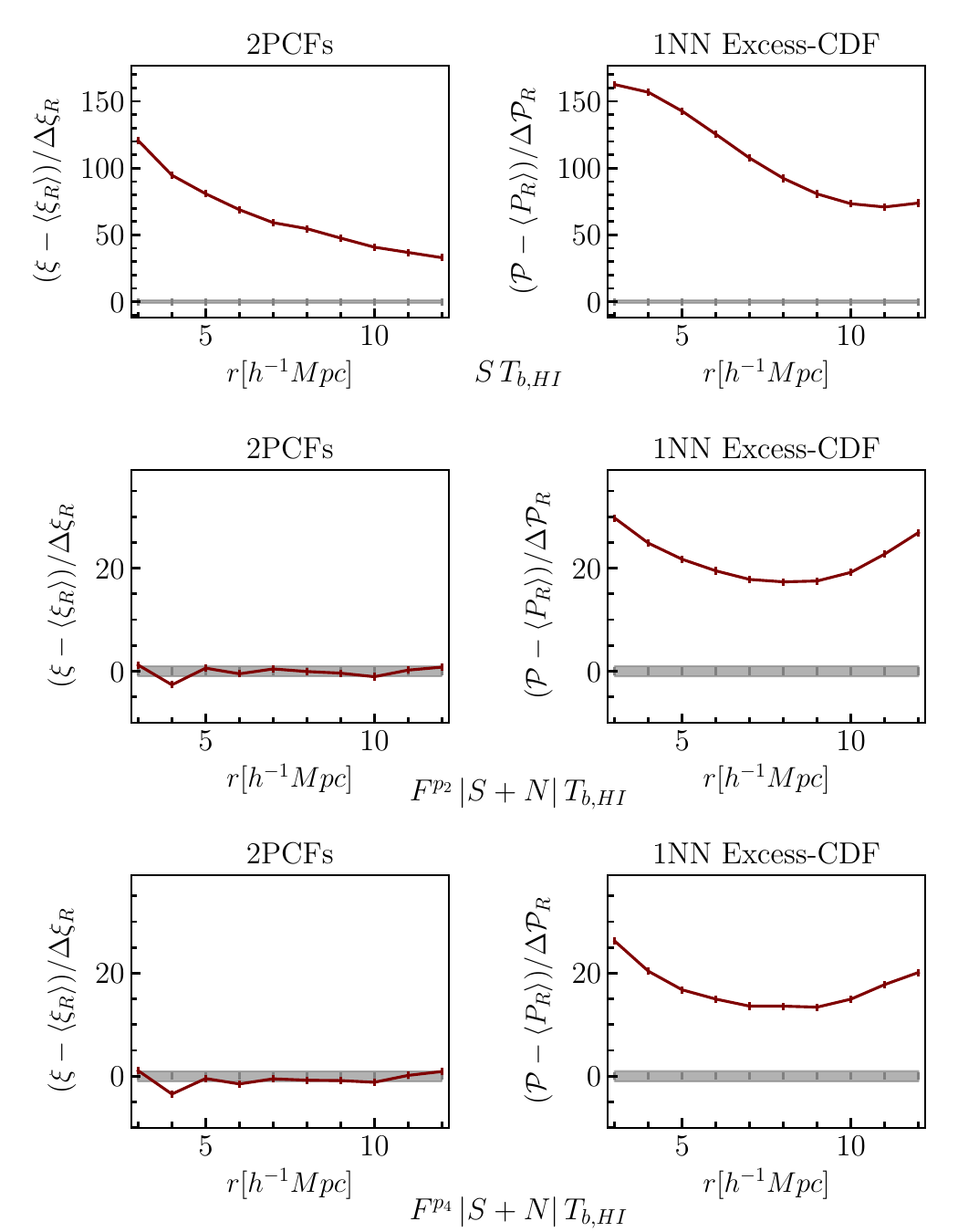}
     \caption{Cross-correlation measurements: Relative SNR per bin (explained in Sec.~\ref{sec:crosscorreln}), plotted for the range $ r=[3,12] \, \hmpc $. The \textit{first row} corresponds to the original simulation field, $ \Stb $. The \textit{second} and \textit{third rows} corresponds to $\tbhi$ field accounted for the thermal noise component for $ 22 $ days of observation time, with foreground filtering applied using $ F^{p_2} $ and $ F^{p_4} $, respectively; both have $ k_{\rm min, \parallel}=0.3 \, \ihmpc $. The \textit{left panel} of each row displays $ 2\pcf $ measurements, while the \textit{right panel} shows the $ 1\nn $ Excess-CDF measurements. The solid maroon line in each panel represents the measured signal, while the grey-shaded band indicates the $1\sigma$ scatter derived from 2000 realizations of randomly chosen points. Thermal noise and foreground removal strongly suppress the $ 2\pcf $ signal and introduce non-physical anti-correlations, while the positively correlated $ 1\nn $ signal remains well above the statistical uncertainty in each bin.}
    \label{fig:relativeSNRcombine}
\end{figure}  

For each radial bin, we compute the relative cross-correlation SNR by subtracting the mean cross-correlation for the randomly sampled points (averaged over 2000 realizations) from the cross-correlation signal for haloes. This difference is then divided by the standard deviation (1$\sigma$ value) of the cross-correlations across the random realizations. 
In Fig.~\ref{fig:relativeSNRcombine}, the solid maroon line in each plot illustrates this relative SNR of the cross-correlation between the $\tbhi$ field and the actual halo positions for a radial range of $r = [3,12] \, \hmpc$. The lower limit on $r$ roughly represents the smallest physical scale probed by CHIME's stacking analysis, while the upper limit corresponds to distances where the nearest neighbor auto-CDF computed in the $k\nn$-field framework with $k=1$, are well-measured and remain well below the largest scale investigated by CHIME \citep{CHIME:2022kvg}.
Our whole analysis is restricted to this particular range. The grey-colored band in the same plots indicates the $1\sigma$ scatter computed for random realizations of points. The first row in the figure corresponds to $\Stb$, whereas the second and third rows correspond to $\FNtbA$ and $\FNtbB$ respectively. Within each row, the left subpanel shows the $2 \pcfs$ measurement while the right subpanel displays the $1\nn$ Excess-CDF measurements.

As anticipated, the per-bin relative SNR for the cross-correlation measurements, as depicted in Fig.~\ref{fig:relativeSNRcombine}, is higher for the $k\nn$-field framework compared to the $2\pcfs$ for all cases of $\tbhi$ shown. Notably, noise contamination and the filtering operation decrease
the signal for both statistics; however, this effect is more pronounced in the $2\pcfs$, which is close to the $1\sigma$ band computed from the randomly sampled points. Additionally, in the $2\pcfs$, there appears to be an anti-correlation (negative $2\pcf$) in a few radial bins. This does not indicate a true physical anti-correlation between haloes and $\hi$ fluctuations; rather, it results from foreground filtering, which produces oscillations in position space due to the filter being applied in Fourier space. 
We will see below that this has a notable effect on the $2\pcf$ SNR after the filtering has been applied. On the other hand, the sharper filter ($F^{p_4}$) decreases the $1\nn$ signal more than the Gaussian filter ($F^{p_2}$), but the signal remains positive in both cases.

It should be noted that while each of the plots in Fig.~\ref{fig:relativeSNRcombine} represents a relative cross-correlation SNR measure, these plots do not depict the complete measure of cross-correlation SNR. This is because off-diagonal terms also appear in the covariance matrix, as will be shown in Sec.~\ref{sec:covarianceR}, and thus need to be accounted for in the full SNR measure. We quantify the cross-correlation SNR using the chi-square analysis presented in Sec.~\ref{sec:chisquare}.

%%%%%%%%%%%%%%%%%%%%%%%%%%%%%%%%%%%%%%%%%%%%%%%%%%
\subsection{Covariance Matrix} 
\label{sec:covarianceR}
%%%%%%%%%%%%%%%%%%%%%%%%%%%%%%%%%%%%%%%%%%%%%%%%%%

\begin{figure}
\centering
  \includegraphics[width=0.95\textwidth]{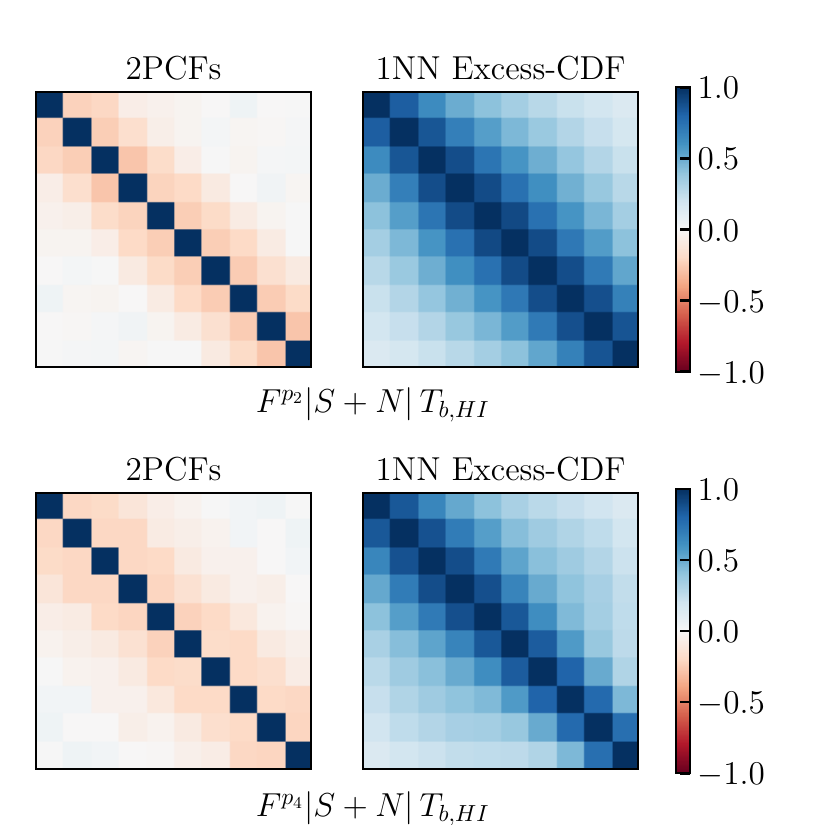}
    \caption{Correlation Matrix: This matrix is computed for contaminated $\tbhi$ fields by cross-correlating the fields with realizations of randomly sampled points (see Sec.~\ref{sec:covarianceR}). The \textit{first row} represents the scenario with the $F^{p_2}$ filter, while the \textit{second row} corresponds to the scenario with the $F^{p_4}$ filter. In each row, the \textit{left panel} shows the measurement of $ 2\pcfs $, whereas the \textit{right panel} presents $ 1\nn $ Excess-CDF measurements computed via the $ k $NN Field framework. The correlation matrix for the $ k $NN Field framework displays positive correlations across different nearby radial scales, stemming from the cumulative nature of the measurements involved. In contrast, the $ 2\pcf $ measurement lacks such positive correlations; instead, very nearby radial bins exhibit anti-correlations due to filtering effects, which are absent in the original $ \hi $ field (see Fig.~\ref{fig:covar_true}). The correlation matrices are not fully diagonal, even for the $ 2\pcfs $ at the involved radial scales. The choice of filter smoothness, with $ F^{p_4} $ being sharper, slightly affects the matrices.}

    \label{fig:covarTBN}
\end{figure}

The covariance matrix captures the variance and covariance of the cross-correlation signal across different radial bins. We compute the covariance matrix from 2000 realizations of randomly sampled points within the IllustrisTNG300 simulation box. In this section, we present the computation of the covariance matrix and its resulting structure. The $(i,j)$ entry of the covariance matrix is given by:
\begin{equation}
    \label{eq:covar}
    C(i,j) = \Bigg\langle \Big (D_{R}^p(i) - \Big \langle D_{R}^p(i) \Big \rangle \Big ) \,\Big (D_{R}^p(j) - \Big \langle D_{R}^p(j) \Big \rangle \Big ) \Bigg\rangle
\end{equation}
Here, $D_{R}^p$ represents the data vector computed for the $p^{\rm th}$ realization of randomly sampled points, with a length of $n_{\rm bin} = 10$, corresponding to the number of radial bins being analyzed. Within the $k\nn$-field framework, $D_{R}^p$ corresponds to the $1\nn$ Excess-CDF measure, while in the context of $2 \pcfs$, it represents the two-point cross-correlation measure. The symbol $\langle \cdots \rangle$ denotes the average across the 2000 random realizations.

The covariance matrix $C$, computed using Eq.~\ref{eq:covar}, has dimensions of $n_{\rm bin} \times n_{\rm bin}$. Its diagonal entries measure the variance of each bin, while the off-diagonal entries measure the correlations between bins. In Fig.~\ref{fig:covarTBN}, we illustrate the structure of the correlation matrix, $C_N$, obtained for the two cases of $\FNtbA$ (first row) and $\FNtbB$ (second row). The elements of the correlation matrix are given by $C_N(i,j) = C(i,j)/\sqrt{C(i,i) \,C(j,j)}$, which normalizes the diagonal entries of $C$ to $1$, providing a reference for the correlation between bins indicated by the off-diagonal entries. The left panels of the figure present measurements of the $2 \pcfs$, while the right panels show measurements of the $1\nn$ Excess-CDF using the $k\nn$-field framework.

For both types of filters, the correlation matrix from the $k\nn$-field framework exhibits significant positive correlations across different radial scales, as evidenced by substantial off-diagonal terms ($>0$), with the $F^{p_2}$ case showing slightly stronger off-diagonal entries than $F^{p_4}$. This aligns with the cumulative nature of the CDF measure in this framework, where fluctuations at smaller radial bins propagate to larger ones, resulting in correlated measurements across bins.
For instance, in the random sampling case, a high upward fluctuation in the Excess-CDF measure at a smaller radial bin can elevate the measurement in the next radial bin compared to a scenario where the previous bin did not exhibit such a fluctuation (similarly for negative fluctuations).

In contrast, the $2\pcf$ measurements do not exhibit such positive correlations; instead, they indicate anti-correlations in nearby radial bins, with the $F^{p_2}$ case showing slightly more anti-correlated nearby bins compared to $F^{p_4}$. The $2\pcf$ technique computes correlations independently for each radial bin, suggesting that the correlations among radial bins are unrelated to our chosen measurement method.
The observed anti-correlation in $2\pcfs$ appears after filtering of the $\tbhi$ field.
As an additional check, we present the structure of the correlation matrix obtained for the original simulated $\tbhi$ field, $\Stb$, in Appendix~\ref{sec:a1}. There, $2\pcfs$ does not show such anti-correlations.

%%%%%%%%%%%%%%%%%%%%%%%%%%%%%%%%%%%%%%%%%%%%%%%%%%
\subsection{Chi-square Analysis}
\label{sec:chisquare}
%%%%%%%%%%%%%%%%%%%%%%%%%%%%%%%%%%%%%%%%%%%%%%%%%%

\begin{figure*}
  \includegraphics[width=0.95\textwidth]{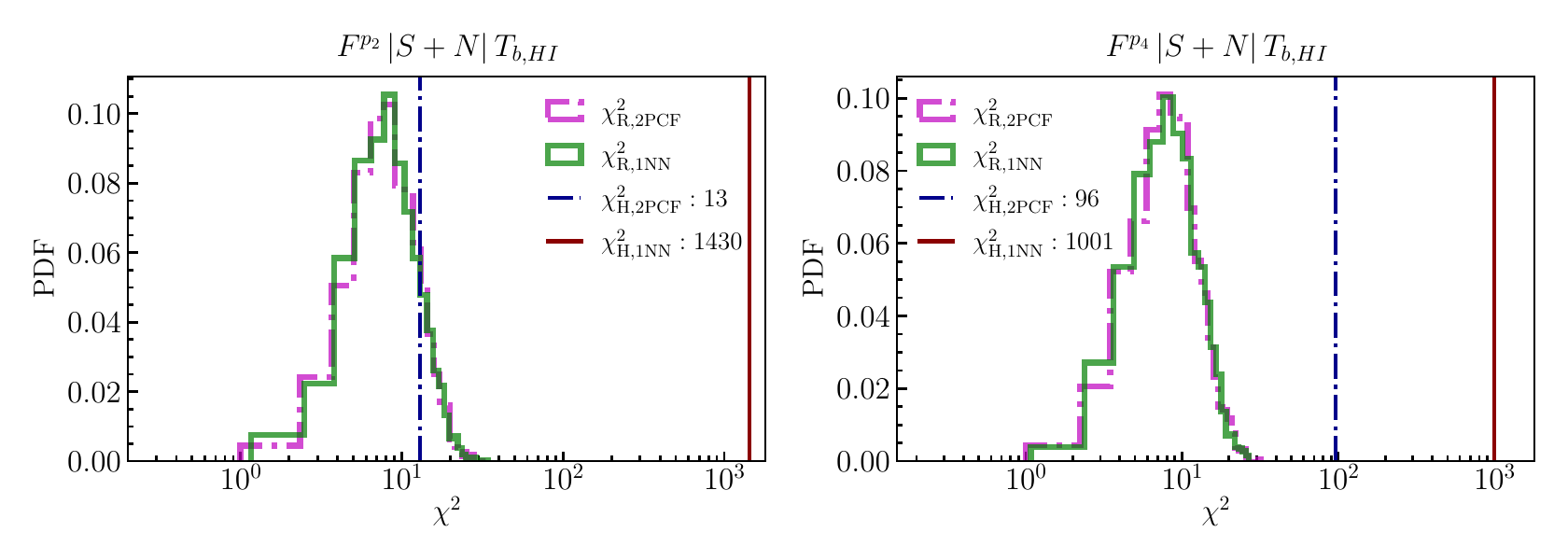}
    \caption{Chi-square analysis results (detailed in Sec.~\ref{sec:chisquare}) for contaminated $\tbhi$ fields: The \textit{left panel} corresponds to the $F^{p_2}$ (smoother) filter case, while the \textit{right panel} corresponds to $F^{p_4}$ (sharper) filter. The dash-dotted navy line represents $ \chi^{2}_{H, 2 \pcf} $, the chi-square value from the $2\pcf$ measurement on actual haloes, with the dash-dotted magenta histogram indicating $ \chi^{2}_{R, 2 \pcf} $ for randomly sampled points (obtained using 2000 realizations). The solid maroon line represents $ \chi^{2}_{H, 1\nn} $ for actual haloes, while the solid green histogram represents $ \chi^{2}_{R, 1\nn} $ values, for the random case. For the $F^{p_2}$ case, the chi-square values are $ \chi^{2}_{H, 2 \pcf} \approx 13 $ (significance of $ 0.8\sigma $, indicating no detection) and $ \chi^{2}_{H, 1\nn} \approx 1430 $ (significance of $ 37\sigma $). Whereas for the $F^{p_4}$ case, $ \chi^{2}_{H, 2\pcf} \approx 96 $ (significance of $ 8\sigma $) and $ \chi^{2}_{H, 1\nn} \approx 1001 $ (significance of $ 31\sigma $). These values reflect the significance with which the no-signal null hypothesis can be rejected. The $2\pcf$ fails to detect a signal for the smoother filter while showing spurious detection significance for the sharper filter (see text). However, the $1\nn$ measurements show a higher detection significance for the smoother filter. These results highlight the superior detectability and potential constraining power of the $ k\nn$-field framework compared to the $ 2\pcf$.}
    \label{fig:chisquareTbFN}
 \end{figure*}

To quantify the $\hi$-galaxy cross-clustering total SNR, we performed a chi-square analysis. The chi-square value is defined as follows:
\begin{equation}
    \chi_{y}^{2} = \sum_{i, j=1}^{\mathrm{n_{bin}}}\Big (D_{y}(i) - \Big \langle D_{R}^p(i) \Big \rangle \Big) \, {C^{-1}(i,j)} \, \Big (D_{y}(j) - \Big \langle D_{R}^p(j) \Big \rangle \Big)
    \label{eq:defnchisq}
\end{equation}

Here, $D_{y}$ denotes the data vector, which in the $k \nn$ case is simply the $1\nn$ Excess-CDF, while in the $2 \pcfs$ case, it represents the two-point cross-correlation measure. When considering realizations of randomly sampled points, $y = \random$, and for the actual positions of haloes, $y = \halo$. The degrees of freedom in this analysis equals the number of radial bins examined, which is 10.

As mentioned previously, our null hypothesis assumes the absence of correlations between haloes and $\hi$ fluctuations. The distribution of chi-square values obtained using realizations of randomly sampled points defines the null hypothesis region, while the chi-square value for actual halo positions quantifies the true cross-clustering signal.
Using Eq.~\ref{eq:defnchisq}, we calculate the chi-square value for the signal case, $\chi_{\halo}^{2}$, and for 2000 sets of random position realizations, each one denoted by $\chi_{\random}^{2}$. The computation of the covariance matrix, $C$, is explained in Sec.~\ref{sec:covarianceR}. The inverse covariance matrix, $C^{-1}$, is adjusted for the finite number of realizations using the Hartlap factor \citep{HartlapFactor}. A significant deviation of $\chi_{\halo}^{2}$ from the null hypothesis region indicates a detection of $\hi$ clustering around the haloes.

The results of the chi-square analysis for both $\FNtbA$ and $\FNtbB$ are presented in Fig.~\ref{fig:chisquareTbFN}. In the figure, the dash-dotted magenta histogram distribution, labeled $\chi^{2}_{\random,2 \pcf}$, represents the null hypothesis region obtained from the $2\pcf$ measurements. The dash-dotted navy line depicts the corresponding signal for actual haloes, $\chi^{2}_{\halo,2\pcf}$. Similarly, the solid green histogram distribution, labeled $\chi^{2}_{\random, 1 \nn}$, represents the null hypothesis region obtained from the $k\nn$-field framework measurement for $k = 1$, with the corresponding signal shown by the solid maroon line, $\chi^{2}_{\halo,1 \nn}$. The distributions of $\chi^{2}_{\random,1 \nn}$ and $\chi^{2}_{\random,2 \pcf}$ closely approximate a chi-square distribution with 10 degrees of freedom.

The chi-square values computed for true halo positions with the $F^{p_2}$ filter are $\chi^{2}_{\halo, 2 \pcf} \approx 13$ and $\chi^{2}_{\halo, 1 \nn} \approx 1430$. In this case, the $2\pcf$ SNR cannot rule out the null hypothesis, indicating no detection. For the sharper filter $F^{p_4}$, we find $\chi^{2}_{\halo, 2 \pcf} \approx 96$ and $\chi^{2}_{\halo, 1 \nn} \approx 1001$. Notably, the sharper filter decreases the chi-square value for $1\nn$ measurements compared to the smoother filter.
Somewhat surprisingly, we also find that the sharper filter \textit{increases} the chi-square for the $2\pcf$ compared to the smoother filter. As mentioned in Sec.~\ref{sec:crosscorreln}, a filter in $k$-space introduces some ringing in the position-space field; if the filter is sufficiently sharp, the associated ringing in the $2\pcf$ can exceed the statistical errors, resulting in a strong detection of the cross-correlation. This ringing can also propagate into the covariance matrix itself, which can likewise alter the detection significance. We explore the size of these effects in Appendix~\ref{sec:b1}, and comment on their implications in Sec.~\ref{sec:varyf&n}.

%%%%%%%%%%%%%%%%%%%%%%%%%%%%%%%%%%%%%%%%%%%%%%%%%%%% Summary table ; Improved FG level and Noise

\begin{table*}
    \centering
    \captionsetup{width=0.85\textwidth} % Centering the caption and setting its width
    \renewcommand{\arraystretch}{1.2} % Adjust this value for more or less space
    \begin{tabular}{|c|c|c|c|c|}
        \hline
        \textbf{$\, \tbhi$ Field} & 
        \multicolumn{2}{c|}{\boldmath$F^{p_2}: 1 - \exp\left(-k^2/k_{\parallel,\rm min}^2 \right)$} & 
        \multicolumn{2}{c|}{\boldmath$F^{p_4}: 1 - \exp\left(-k^4/k_{\parallel,\rm min}^4 \right)$} \\
        \cline{2-5} % Line below the multicolumn section
        & \textbf{$\sigma_{2\pcf}$} & \textbf{$\sigma_{1\nn}$} & 
          \textbf{$\sigma_{2\pcf}$} & \textbf{$\sigma_{1\nn}$} \\
          \hline
            $\FeNdtb$ &  75 & 173 &  80  & 173  \\
            $\FdNdtb$ &  60 & 176 &  58  & 167  \\
            $\FcNdtb$ &  43 & 181 &  44  & 167  \\
            $\FbNdtb$ &  27 & 177 &  33  & 163  \\
            $\FaNdtb$ &  18 & 181 &  41  & 162  \\ 
            $\FNdtb$  &  15 & 180 &  57  & 163  \\
            \hline
            $\FNctb$ &  10  & 128 & 38  & 114  \\
            $\FNbtb$ &  9   & 98  & 29  & 85   \\
            $\FNatb$ &  2.9 & 57  & 16  & 49  \\
            $\FNtb$  &  0.8 & 37  & 8   & 31  \\
            \hline
    \end{tabular}

    \caption{Detection significance for the $k\nn$-field framework ($\sigma_{1\nn}$) and the $2\pcfs$ ($\sigma_{2\pcf}$) for $\tbhi$ fields across varying noise levels, parameterized by integration time $t_{\rm obs}$, and foreground cleaning levels, parameterized by $k_{\rm min,\parallel}$. The $\tbhi$ field notation follows the format $F^{u}_{k_{\rm min, \parallel}}|S+N_{t_{\rm obs}}|$, where $u$ represents the filter type: $p_2$ for the smoother (Gaussian) filter and $p_4$ for the sharper filter. Refer to Sec.~\ref{sec:varyf&n} for a detailed discussion of these results.}

    \label{tab:Summary}
\end{table*}

To express the chi-square value in terms of the effective number of sigmas ($\sigma$), which is commonly used to quantify signal detection significance, we employ the following relation \citep{CHIME:2022kvg}:
\begin{equation}
    \sigma = \Phi^{-1} \left[1 - \int_{s}^{\infty} \chi_{\nu}^{2}(x) \,dx\right],
    \label{eq:sigma}
\end{equation}
where $\Phi^{-1}$ denotes the inverse CDF of the standard normal distribution, $s$ is the chi-square value for which the corresponding $\sigma$ is being computed, and $\nu$ represents the degrees of freedom, which in our case is 10. This conversion relation is a rough approximation and becomes more accurate when the chi-square distribution is well approximated by a Gaussian distribution. 

Using Eq.~\ref{eq:sigma}, we find for $\FNtbA$, $\sigma_{2\pcf} = 0.8$ and $\sigma_{1\nn} = 37$, and for $\FNtbB$, we obtain $\sigma_{2\pcf} = 8$ and $\sigma_{1\nn} = 31$. These numbers indicate that the $k\nn$-field framework, unlike $2\pcfs$, yields a significantly high cross-clustering signal compared to the statistical noise, even in the presence of significant foreground filtering. This underscores its potential for enhanced constraining power compared to $2\pcfs$. These findings are robust across different filter choices, as we discuss in the next subsection.

%%%%%%%%%%%%%%%%%%%%%%%%%%%%%%%%%%%%%%%%%%%%%%%%%%
\subsection{Improved Levels of Foreground Cleaning and Thermal Noise}
\label{sec:varyf&n}
%%%%%%%%%%%%%%%%%%%%%%%%%%%%%%%%%%%%%%%%%%%%%%%%%%

In this subsection, we explore how the results of Sec.~\ref{sec:chisquare} vary when we alter the assumed thermal noise or aggressiveness of 21 cm foreground cleaning, respectively parameterized by integration time $t_{\rm obs}$ and line-of-sight filtering threshold $k_{\rm min, \parallel}$. Improvements on both fronts are achievable with advancements in instrumental design, foreground cleaning techniques, and data processing.

We use the notation $\Ntgeneral$ to denote the $\tbhi$ field with noise corresponding to the integration time $t_{\rm obs}$, and $\FNtgeneral$ to denote the case with integration time $t_{\rm obs}$ and the foreground cleaning corresponding to the given $k_{\rm min, \parallel}$ value.
We consider integration times of 2 months, 6 months, 1 year, and 4 years, with $k_{\rm min,\parallel}$ values spaced regularly between $0.05\,\ihmpc$ and $0.3\,\ihmpc$.
Using our pipeline, we examine the cross-correlation of these fields with the set of 4000 most massive haloes. Table~\ref{tab:Summary} shows the final results, including chi-square values and their related effective sigma levels. 

First, we focus on the results from traditional $2\pcf$ measurements. For both $F^{p_2}$ (smoother) and $F^{p_4}$ (sharper) filters, we observe a gradual increase in total SNR as the noise level decreases (indicated by longer $t_{\rm obs}$) while maintaining a constant filtering level ($k_{\rm min, \parallel} = 0.3\, \ihmpc$). When we fix the noise level to the most optimistic scenario, corresponding to $t_{\rm obs} = 4 \,\rm years $, and decrease the level of filtering (by lowering $k_{\rm min, \parallel}$), for smoother filter we see an expected increase in total SNR. In contrast, for the sharper filter, we observe a non-monotonic trend in SNR, when varying $k_{\rm min, \parallel}$; specifically, for $k_{\rm min, \parallel} \geq 0.2 \,\ihmpc$, the total SNR tends to increase.
We demonstrate in Appendix~\ref{sec:b1} that this is due to the position-space ringing effect mentioned earlier: for the higher values of $k_{\rm min, \parallel}$ and the sharp filter, the $2\pcf$ becomes negative over a certain range of scales (see Fig.~\ref{fig:b1}).\footnote{We have also reproduced this phenomenon by applying the foreground filter to a toy model with a power-law power spectrum.}
As indicated by Eq.~\ref{eq:defnchisq}, the chi-square value (and thus the sigma value) quantifies how much the observed signal deviates from the null hypothesis of zero cross-correlation, accounting for both positive and negative correlations. Therefore, sufficiently strong ringing can act to increase the chi-square value.
Another important observation from Table \ref{tab:Summary}, is that for the sharper filter, the $2\pcf$ SNR is generally higher than that for the smoother filter. Thus, we conclude that $2\pcfs$ are highly sensitive to both the sharpness and intensity of the applied filtering.

Now, we discuss the results obtained from the $k\nn$-field framework measurements. We observe no anti-correlations between haloes and $\tbhi$ field (the $1\nn$ Excess-CDF consistently remains positive) due to the field's filtering, regardless of the sharpness or level of filtering, unlike the case of $2\pcfs$ (see Fig.~\ref{fig:b1}).
As shown in Table~\ref{tab:Summary}, the $1\nn$ Excess-CDF total SNR is generally higher for a smoother filter like $F^{p_2}$ compared to $F^{p_4}$, which is intuitive because the smoother filter causes lower attenuation of lower-$k_\parallel$ information. We also reanalyzed the cross-clustering SNR for $F^{p_2}$ case, using another set of 2000 such realizations of randomly sampled points, which yielded similar comparisons between the $F^{p_2}$ and $F^{p_4}$ cases.
For both types of filters, the $\sigma$ values are stable to within $\sim$5\% as $k_{\rm min, \parallel}$ increases within the range tested.
This indicates that moderate improvements in foreground filtering will not significantly affect the SNR with which the $1\nn$ Excess-CDF can be measured.

The non-monotonic behavior of $2\pcf$ SNR for a sharper filter like $F^{p_4}$ indicates that, with sufficiently aggressive foreground filtering, the computed SNR does not directly reflect the cosmological information recoverable from $2\pcf$ measurements: we expect the information content to decrease as $k_{\rm min, \parallel}$ increases, yet the SNR can \textit{increase} with increasing $k_{\rm min, \parallel}$ in a certain regime. Quantifying this information content is beyond the scope of this work, but will be essential to fully compare the utility of the $2\pcf$ and $k\nn$-field framework. Therefore, the ratios of $k\nn$ and $2\pcf$ SNR values in Table~\ref{tab:Summary} for the $F^{p_4}$ case should be interpreted as \textit{lower bounds} on the improvement in constraining power that can be obtained using $k\nn$ measurements instead of the $2\pcf$. 
On the other hand, for a smoother filter like $F^{p_2}$, the ringing effects are not strong enough to affect the expected trend of SNR with $k_{\rm min, \parallel}$. Its comparison with the $1\nn$ measurements further establishes the significance of the $k\nn$-field framework, which is also not very sensitive to the choice of filtering, likely because ringing effects are averaged over when spherically-averaged quantities are computed for the $k\nn$ measurements.
As in the fixed-$k_{\rm min, \parallel}$ case, for the $F^{p_4}$ case the ratio of SNR between $k\nn$ and $2\pcf$ measurements is consistently $\gtrsim 3$, whereas for the $F^{p_2}$ case the ratio is always $\gtrsim 10$. This suggests that the higher detectability of the $k\nn$-field cross-correlations is robust to varying assumptions about the effectiveness of 21 cm foreground filtering.

%%%%%%%%%%%%%%%%%%%%%%%%%%%%%%%%%%%%%%%%%%%%%%%%%%
\section{SUMMARY AND OUTLOOK}
\label{sec:summary}
%%%%%%%%%%%%%%%%%%%%%%%%%%%%%%%%%%%%%%%%%%%%%%%%%%

Our primary goal during this work was to develop tools to enhance the $\hi$-galaxy cross-clustering signal-to-noise ratio (SNR) by utilizing higher-order cross-correlations through the $k\nn$-field framework. In contrast, the traditional two-point correlation statistics are limited to second-order cross-correlations. 
To achieve this, we developed cross-correlation analysis pipelines using techniques from both the $k\nn$-field framework and the two-point statistics. Our study employed the IllustrisTNG300 simulation dataset in real space, which includes the simulated 21 cm brightness temperature field ($\Stb$) and halo catalogs (mock galaxies), both at a redshift of $z = 1$. 
We computed the detection significance (in the form of chi-square values, which we converted into the more familiar ``number of sigmas'') for the $k\nn$-field framework (specifically, the $1\nn$ Excess-CDF) alongside the standard two-point cross-correlation function ($2\pcf$), under varying assumptions for 21 cm thermal noise and foreground filtering. This allowed us to compare the detectability of the two methods across different cases. 
The comparison between using $k=1$ and combining higher $k$ values (specifically, $k=2$) within the radial range $r = [3,12] \, \hmpc$ showed comparable significance in the cross-clustering detection. As a result, we presented findings based solely on $k=1$. However, incorporating higher $k$ values could potentially extend the analysis to larger scales and break parameter degeneracies that may exist in the $k=1$ case.

The main conclusions from our analysis are as follows:
\begin{itemize}
%%%%
\item Assuming an integration time of $t_{\rm obs}=22$ days and foreground filtering that removes Fourier modes with $k_\parallel<0.3\,\ihmpc$ (roughly matching the properties of the CHIME-eBOSS cross-correlation analysis from \citet{CHIME:2022kvg}), we find that the SNR for $1\nn$ Excess-CDF is significantly larger than for the $2\pcf$.
%%%%
\item We consider two forms of foreground filters: a Gaussian filter that suppresses modes with $k_\parallel \lesssim k_{{\rm min},\parallel}$, defined by $F^{p_2}(k_\parallel) = 1 - \exp(-k_\parallel^2 / 2k_{{\rm min},\parallel}^2)$, and an alternative filter that imposes a sharper boundary, $F^{p_4}(k_\parallel) = 1 - \exp(-k_\parallel^4 / 2k_{{\rm min},\parallel}^4)$. 
For fixed $k_{{\rm min},\parallel}$, we find that the choice of filter has a minor impact on the SNR of the $1\nn$ Excess-CDF ($37\sigma$ for $F^{p_2}$ and $31\sigma$ for $F^{p_4}$), but this choice has a strong impact on the $2\pcf$ SNR, which changes from $0.8\sigma$ for $F^{p_2}$ to $8\sigma$ for $F^{p_4}$. 
We demonstrate that this arises from position-space ringing caused by applying the filter to the brightness temperature field in Fourier space: for a sufficiently sharp filter, the ringing is large enough to actually \textit{increase} the detectability of the $2\pcf$ for the filtered field.
%%%%
\item For fixed $k_{{\rm min},\parallel}$, the SNR increases as the assumed thermal noise is decreased (i.e., the assumed integration time $t_{\rm obs}$ is increased).
%%%%
\item For fixed $t_{\rm obs}$, the SNR for the $1\nn$ Excess-CDF is stable (to within $5\%$) with respect to $k_{{\rm min},\parallel}$, when $k_{{\rm min},\parallel}$ is varied between $0.05\,\ihmpc$ and $0.3\,\ihmpc$. With the Gaussian filter, the $2\pcf$ SNR decreases with increasing $k_{{\rm min},\parallel}$, while for the sharper filter, the $2\pcf$ SNR is non-monotonic with $k_{{\rm min},\parallel}$: it decreases as $k_{{\rm min},\parallel}$ increases up to $0.2\,\ihmpc$, and then increases as $k_{{\rm min},\parallel}$ is raised higher than that. We attribute this behavior to the same position-space ringing mentioned above and provide plots that demonstrate the effect of ringing on the $2\pcf$ signal as a function of $k_{{\rm min},\parallel}$.
\end{itemize}

The above conclusions raise questions about the relationship between the detectability of the 21 cm-galaxy $2\pcf$ and the amount of cosmological information retrievable from this measurement in the presence of foreground filtering. Future work should investigate the cosmological constraining power of the filtered $2\pcf$ to better interpret the SNR values obtained from simulations. Nonetheless, a clear general conclusion emerges: the $1\nn$ Excess-CDF achieves a much higher SNR than the $2\pcf$ across a broad range of assumptions about thermal noise and foreground filtering. 

To get a rough sense of the constraining power from the obtained SNR values, consider an idealized case where the scale dependence of the signal is fully known, leaving the overall amplitude as the only free parameter. In this case, the SNR directly translates into the precision of the fitted amplitude. For instance, an SNR of 37 for the $1\nn$ Excess-CDF implies an amplitude precision of $100\% \times 1/37 = 2.7\%$, whereas the $2\pcf$ SNR of 0.8 corresponds to a much poorer precision. 
This argument is commonly used to develop intuition for precise measurements in CMB lensing studies \citep[see, e.g.,][]{Carron2022, ACT2024, Shaikh2024}.  

Despite the promising results in this paper, one may question how accurately we can model the abundance and clustering of both neutral hydrogen and galaxies on non-linear scales needed to exploit the $k\nn$ techniques employed in this work fully. While both tracers can be challenging to model in the non-linear regime, the community is constantly developing tools to address these issues. In particular, state-of-the-art hydrodynamic simulations \citep[see, e.g.,][]{CAMELS_presentation} that vary in both cosmology and astrophysics are designed to provide accurate theoretical predictions for the abundance and clustering of galaxies, $\hi$, and many other properties while accounting for baryonic uncertainties on small scales.
The $\hi$-galaxy connection itself is influenced by baryonic processes \citep[see, e.g.,][]{Osinga2024}, further complicating these predictions. Semi-analytical models can also be used for such endeavors given the fact that in the post-reionization era, most of the $\hi$ resides within galaxies. These tools can be used in conjunction with the $k\nn$-field framework presented in this work to perform parameter inference while incorporating our incomplete knowledge of the impact of astrophysical processes on the clustering of $\hi$ and galaxies.

Therefore, applying the $k\nn$-field framework to observational 21 cm data from ongoing $\hi$ intensity mapping surveys like CHIME, along with galaxy data from optical surveys such as eBOSS and DESI, holds significant promise.
The $k\nn$-field framework captures a wealth of information that two-point statistics overlook, primarily due to their limitation in capturing only Gaussian information. Therefore, once we have the $\hi$ detection via the $k\nn$-field framework, it is vital to further develop techniques to model the signal. One approach was explored in \citet{Banerjee2022}, but new techniques need to be explored for the specific context discussed in this paper. 

There are several other avenues to explore in future work:
\begin{itemize}
%%%
\item As discussed above, an important question that arises is how the cosmological constraining power of the $1\nn$ Excess-CDF compares to that of the $2\pcf$, particularly under various assumptions about 21 cm foreground filtering. Additionally, it is essential to compare the $1\nn$ Excess-CDF with the 21 cm galaxy cross-power spectrum, as applying a filter that erases modes with $k_\parallel < k_{\rm min,\parallel}$ will have more straightforward effects on the cross-power spectrum than on the $2\pcf$.
%%%
\item Investigating a joint analysis of the standard 2-point cross-correlation and the $1\nn$ Excess-CDF would clarify the amount of information that is shared between the two probes, as well as the information that is captured by $k\nn$ measurements but not by the $2\pcf$.
%%%
\item In auto-clustering studies, combining measurements with $k > 1$ enhances the constraining power \citep{kNN}. In our analysis of $\hi$-galaxy cross-correlations, we recognize the potential benefits of incorporating higher $k$ values to increase the information content. However, the extent of the additional information provided by $k > 2$ remains to be explored.
%%%
\item While previous studies have addressed how to handle masked data in other observational datasets \citep{Kaustubh2024, Wang2022} for $k\nn$ statistics, it is particularly important in the context of this work to consider the handling of masked data in $\hi$ and galaxy datasets when measuring the $k\nn$-field framework.
%%%
\item In this work, we considered real-space data to establish a proof of concept. Further investigation into how redshift-space distortions affect the results will be crucial when analyzing real observational data.
%%%
\item \citet{Coulton2024} recently demonstrated the utility of the $k\nn$-field framework for detecting the imprints of primordial non-Gaussianity in a general context. Investigating the applicability of this framework to $\hi$-galaxy cross-correlations is an important area for further research.
\end{itemize}

%%%%%%%%%%%%%%%%%%%%%%%%%%%%%%%%%%%%%%%%%%%%%%%%%%%%%
\section*{Acknowledgements}
%%%%%%%%%%%%%%%%%%%%%%%%%%%%%%%%%%%%%%%%%%%%%%%%%%%%%

EC and AB would like to thank their research group members, especially Kaustubh Rajesh Gupta, for their regular discussions on this work, which proved to be immensely helpful. EC acknowledges the Ministry of Education, Government of India, and the Indian Institute of Science Education and Research, Pune, for the support of the Prime Minister Research Fellowship (PMRF). AB's work was partially supported by the Startup Research Grant (SRG/2023/000378) from the Science and Engineering Research Board (SERB), India. We also acknowledge the PARAM Brahma Facility under the National Supercomputing Mission, Government of India, at the Indian Institute of Science Education and Research, Pune, for providing the computing resources for this work.
This material is based upon work supported by the U.S.\ Department of Energy, Office of Science, Office of High Energy Physics under Award Number DE-SC0024309.

%%%%%%%%%%%%%%%%%%%%%%%%%%%%%%%%%%%%%%%%%%%%%%%%%%%%%
\section*{Data Availability}
%%%%%%%%%%%%%%%%%%%%%%%%%%%%%%%%%%%%%%%%%%%%%%%%%%%%%
The $\hi$ and halo datasets utilized in this study are publicly available at \url{https://www.tng-project.org/data/}. The data generated during this study can be made available upon reasonable request.



\bibliographystyle{mnras}
\bibliography{reference} % if your bibtex file is called reference.bib

% this method is tedious and prone to error if you have lots of references
%\begin{thebibliography}{99}
%\end{thebibliography}

%%%%%%%%%%%%%%%%%%%%%%%%%%%%%%%%%%%%%%%%%%%%%%%%%%

%%%%%%%%%%%%%%%%% APPENDICES %%%%%%%%%%%%%%%%%%%%%

\appendix

%%%%%%%%%%%%%%%%%%%%%%%%%%%%%%%%%%%%%%%%%%%%%%%%%%%%%
\section{$\hi$ Clustering in $\Stb$ } 
\label{sec:a1}
%%%%%%%%%%%%%%%%%%%%%%%%%%%%%%%%%%%%%%%%%%%%%%%%%%%%%

\begin{figure}
\centering
\includegraphics[width=0.9\textwidth]{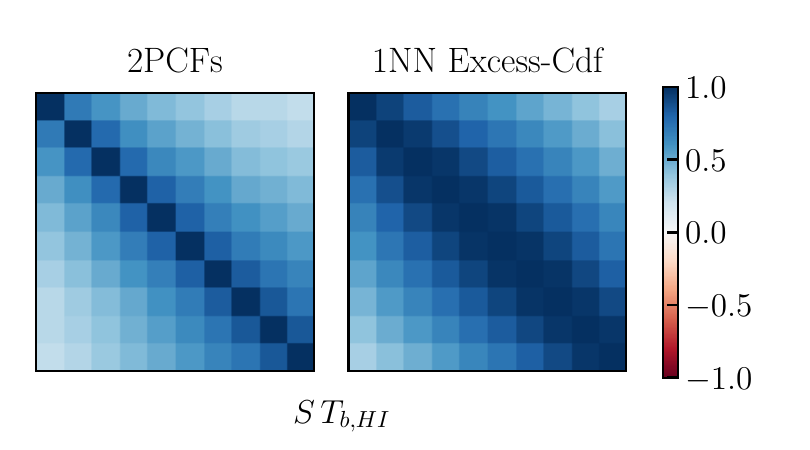}
   \caption{$\Stb$ Correlation Matrix: The matrix represents the correlations between different bins, computed using 2000 realizations of randomly sampled points cross-correlated with $\Stb$. The \textit{left panel} corresponds to the $2\pcf$ measurements, while the \textit{right panel} corresponds to the $1\nn$ Excess-CDF measurements obtained via the $k\nn$-field framework. The $k\nn$-field framework exhibits significant positive correlations among different radial scales, which, as discussed in Sec.~\ref{sec:covarianceR}, are inherent to the measurements in this framework. In contrast, the $2\pcf$ measurements show relatively smaller correlations between nearby radial bins, which are not intrinsic to the measurement technique of the $2\pcf$ itself.}
    \label{fig:covar_true}
\end{figure}

We used the original simulated datasets from IllustrisTNG300, consisting of the selected 4000 most massive haloes and the simulated 21 cm brightness temperature field ($\Stb$), to validate the cross-correlation analysis pipeline for both the $k\nn$-field framework and the $2 \pcfs$. Here, we present the validation results.

We computed the covariance matrix using randomly sampled points, as detailed in Sec.~\ref{sec:covarianceR}. For the $\Stb$ field shown in Fig.\ref{fig:STb}, the structure of its correlation matrix is illustrated in Fig.~\ref{fig:covar_true}. The left panel in this figure shows the $2\pcf$ measurements, while the right panel displays the $1\nn$ Excess-CDF measurements within the $k\nn$-field framework. The $k\nn$-field framework demonstrates significant positive correlations among different radial bins, which, as previously mentioned, stem partly from the cumulative nature of the measurements within this framework. In contrast, the correlation matrix structure for the $2\pcf$ reveals positive correlations primarily between very nearby radial bins. These correlations arise from the characteristics of the field and scales considered, rather than being intrinsic to the method of computing the $2\pcf$.

\begin{figure} 
\includegraphics[width=0.98\textwidth]{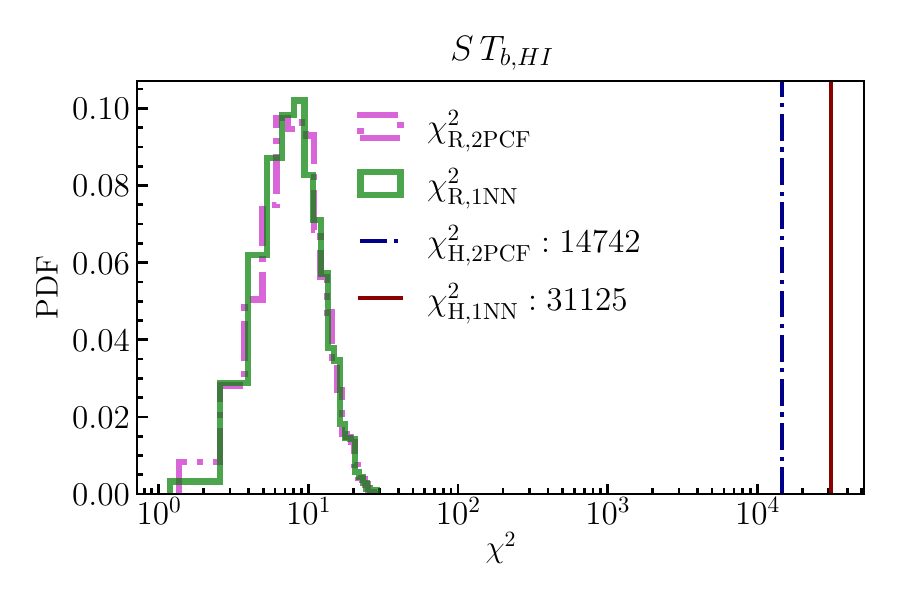}
    \caption{$\Stb$ Chi-Square Analysis Results: The colored representations are consistent with those in Fig.~\ref{fig:chisquareTbFN}. The chi-square values are approximately $\chi^{2}_{H, 2 \pcf} \approx 14742$, corresponding to a significance of approximately $121 \sigma$, while $\chi^{2}_{H, 1 \nn} \approx 31125$ corresponds to a significance of approximately $176 \sigma$. The higher chi-square and effective sigma values from the $1\nn$ measurements, compared to the $2\pcf$, demonstrate that the $k\nn$-field framework is significantly more effective in capturing the cross-clustering of $\hi$ around haloes.}
    \label{fig:chisqtrue}
\end{figure}

In Sec.~\ref{sec:chisquare}, we detailed the chi-square analysis used to quantify the signal-to-noise ratio of cross-correlation measurements and to test the $\hi$ clustering hypothesis. This analysis was initially conducted using $\Stb$ for validation and Fig.~\ref{fig:chisqtrue} presents the corresponding results. The dash-dotted navy line in the figure represents the chi-square value for haloes obtained using the $2\pcf$, denoted as $\chi^{2}_{H, 2 \pcf} \approx 14742$, which corresponds to a significance of approximately $121 \sigma$. For comparison, the chi-square values obtained using realizations of randomly sampled points (uncorrelated case) are shown in the magenta dash-dotted histogram, labeled $\chi^{2}_{R, 2 \pcf}$ (similar to the representations in Fig.~\ref{fig:chisquareTbFN}).
The solid maroon line represents the chi-square value obtained for haloes using the $1\nn$ Excess-CDF measurement, denoted as $\chi^{2}_{H, 1 \nn} \approx 31125$, corresponding to a significance of approximately $176 \sigma$. For randomly sampled points, the chi-square distribution is shown in the solid green histogram, labeled as $\chi^{2}_{R, 1 \nn}$.

These findings demonstrate that both statistics effectively capture the cross-clustering between the haloes and $\hi$ fluctuations present in $\Stb$. However, the $k\nn$-field framework exhibits greater constraining power, maintaining its significance even with contaminated $\tb$ fields (as shown in Sec.~\ref{sec:Results}). This underscores the robustness and effectiveness of the $k\nn$-field framework in such scenarios.

%%%%%%%%%%%%%%%%%%%%%%%%%%%%%%%%%%%%%%%%%%%%%%%%%%%%%
\section{Impact of  Foreground Filtering}
\label{sec:b1}
%%%%%%%%%%%%%%%%%%%%%%%%%%%%%%%%%%%%%%%%%%%%%%%%%%%%%

\begin{figure*} 
\centering
\includegraphics[width=0.85\textwidth]{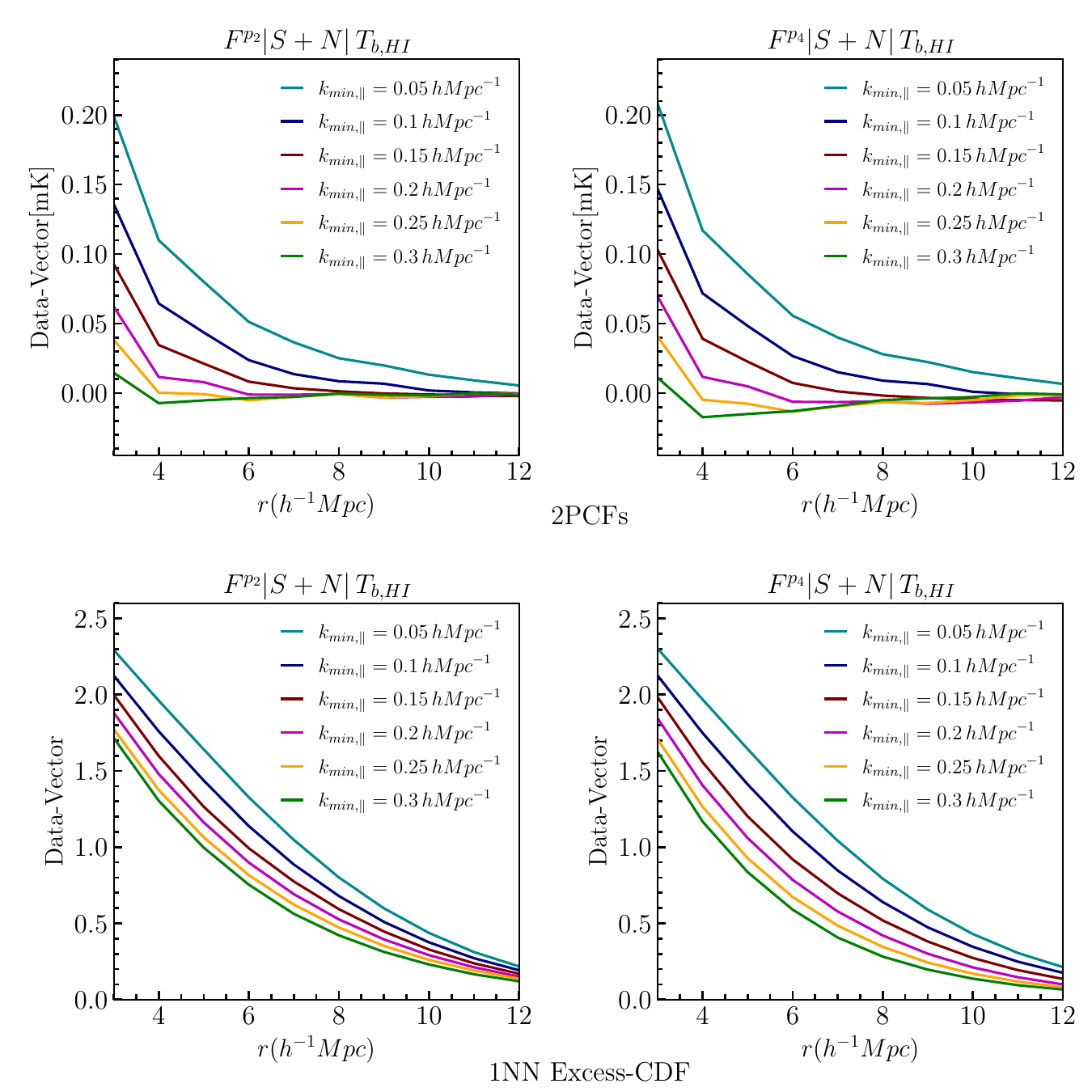}
    \caption{Cross-correlation data vectors (computed as described in Sec.~\ref{sec:Results}) are shown for the case of contaminated $\tbhi$ fields with a fixed $t_{\rm obs} = 4$ years and varying ${k_{\rm min, \parallel}}$ values. The \textit{first row} presents the $2\pcf$ results, where anti-correlations (negative data vectors) are evident in some radial bins for fields with more aggressive filtering, particularly enhanced with a sharper filter like $F^{p_4}$ (\textit{right panel}) in comparison to $F^{p_2}$ (\textit{left panel}). The \textit{second row} displays the $1\nn$ Excess-CDF cross-correlation data vectors. Unlike the $2\pcfs$, rigorous filtering does not introduce anti-correlations in the $k\nn$-field framework. Instead, we observe an expected decrease in the signal as the filtering level (${k_{\rm min, \parallel}}$) increases. Additionally, for the sharper filter $F^{p_4}$, the $1\nn$ signal measure decreases slightly compared to $F^{p_2}$, as the ${k_{\rm min, \parallel}}$ is increased.}
    \label{fig:b1}
\end{figure*}

As presented in Table~\ref{tab:Summary}, which summarizes the total SNR and detection significance results for various choices of foreground filters (${k_{\rm min, \parallel}}$) and thermal noise levels ($t_{\rm obs}$), we observe that for a sharper $F^{p_4}$ filter choice, with a fixed thermal noise level (corresponding to $t_{\rm obs} = 4$ years, the most optimistic scenario) and increasing ${k_{\rm min, \parallel}}$ (indicating more aggressive foreground cleaning), the trend in the total $2\pcf$ SNR is not monotonic. Specifically, for ${k_{\rm min, \parallel}} \geq 0.2 \,\ihmpc$, the SNR increases instead of decreasing (as one would naively expect). We have explained the factors contributing to this behavior in Sec.~\ref{sec:varyf&n} and present further relevant plots in this section.

The first row of Fig.~\ref{fig:b1} presents the $2\pcf$ data vectors, indicating that a sharper filter, such as $F^{p_4}$ (right panel), enhances anti-correlations at higher ${k_{\rm min, \parallel}}$ values compared to the smoother filter $F^{p_2}$ (left panel). Additionally, positive correlations at lower ${k_{\rm min, \parallel}}$ values are also slightly enhanced for $F^{p_4}$. We attribute this behavior to the ringing effects caused by filtering the field in Fourier space, which is more pronounced with a sharper filter.
The second row of the same figure displays the data vectors obtained through the $k\nn$-field framework ($k=1$). This comparison shows that as the levels of foreground filtering increase, the $1\nn$ Excess-CDFs gradually decrease while consistently exhibiting positive correlations. Unlike the $2\pcf$s, this framework does not introduce anti-correlations due to filtering. Furthermore, the $1\nn$ data vectors exhibit a slight decrease with an increase in the sharpness of the filter, particularly at higher ${k_{\rm min, \parallel}}$ values.

\begin{figure*}
\centering
\includegraphics[width=0.85\textwidth]{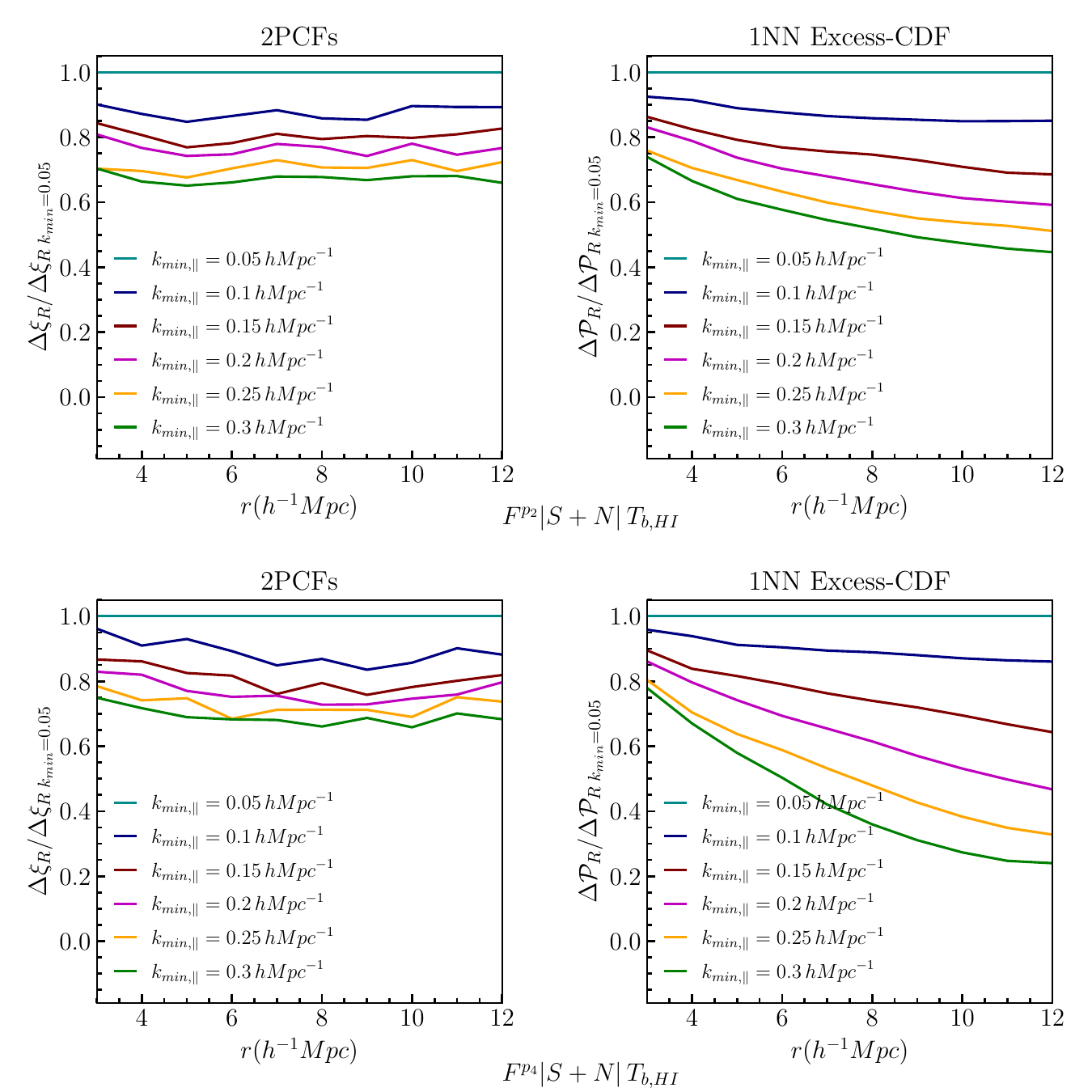}
    \caption{The standard deviation per radial bin in cross-correlations across realizations of randomly sampled points is shown for contaminated $\tbhi$ fields, with fixed thermal noise and varying filtering levels determined by ${k_{\rm min, \parallel}}$ values. The \textit{first row} corresponds to the case $F^{p_2}$ filtering, while the \textit{second row} refers to the $F^{p_4}$ case. The different cases of $\tbhi$ fields presented are consistent with those in Fig.~\ref{fig:b1}, ensuring uniform color coding. The ratios are plotted relative to the ${k_{\rm min, \parallel}} = 0.05\, \ihmpc$ value. The \textit{left panel} displays the $2\pcf$ measurements, while the \textit{right panel} shows the $1\nn$ Excess-CDF measurements. Notably, the noise level decreases significantly more in the $k\nn$-field framework for both types of filters, and this decrease is not uniform across scales.}
    \label{fig:b2}
\end{figure*}

Fig.~\ref{fig:b2} illustrates the behavior of statistical uncertainties (per radial bin) in cross-correlation for various contaminated $\tbhi$ fields subjected to different levels of filtering, all under a fixed instrumental noise level ($t_{\rm obs}=4$ years). We have plotted the ratio of the standard deviation in the contaminated field-random points cross-correlations relative to the case where ${k_{\rm min, \parallel}} = 0.05 \,\ihmpc$. The left panel corresponds to the $2\pcf$ measurements, while the right panel presents the $1\nn$ Excess-CDF measurements, with the first row depicting the $F^{p_2}$ case and the second row depicting the $F^{p_4}$ case. The color coding for different ${k_{\rm min, \parallel}}$ values is consistent with that used in Fig.~\ref{fig:b1}. Notably, noise levels decrease with increased aggressiveness of filtering, with a more pronounced reduction observed in the $k\nn$-field framework. Although the covariance matrix is not fully diagonal, indicating that variance is not a complete measure of noise, the figure effectively illustrates this trend for individual radial bins.

% Don't change these lines
\bsp	% typesetting comment
\label{lastpage}
\end{document}